# US Adaptive Optics Roadmap to Achieve Astro2020

## Committee

| | |
|---|---|
| Julian Christou[1] (Chair) | Gemini Observatory, NOIRLab |
| Mark Chun | University of Hawai'i |
| Richard Dekany | California Institute of Technology |
| Philip Hinz | University of California, Santa Cruz |
| Jessica Lu | University of California, Berkeley |
| Jared Males | University of Arizona |
| Peter Wizinowich | W. M. Keck Observatory |

[1] julian.christou@noirlab.edu

## Workshop Participants

| | |
|---|---|
| Mark Ammons | Lawrence Livermore National Lab. |
| David Anderson | TMT International Observatory |
| Christoph Baranec | University of Hawai'i |
| Rachel Beaton | Space Telescope Science Institute |
| Rebecca Bernstein | GMT, Carnegie Observatories |
| Michael Bottom | University of Hawai'i |
| Corinne Boyer | TMT International Observatory |
| Christophe Clergeron | Gemini Observatory, NOIRLab |
| Jonathan Crass | The Ohio State University |
| Ruben Diaz | Gemini Observatory, NOIRLab |
| Shawn Domagal-Goldman | Goddard Space Flight Center, NASA |
| Justin Finke | Naval Research Laboratory |
| Olivier Guyon | Subaru Telescope / NAOJ, U. Arizona |
| Rebecca Jensen-Clem | University of California, Santa Cruz |
| Masen Lamb | Gemini Observatory, NOIRLab |
| Michael Liu | University of Hawai'i |
| Christopher Martin | California Institute of Technology |
| Dimitri Mawet | California Institute of Technology |
| Oscar Montoya | University of Arizona |
| Zoran Ninkov | National Science Foundation |
| Alison Peck | National Science Foundation |
| Saavidra Perera | University of California, San Diego |
| Eliad Peretz | Goddard Space Flight Center, NASA |
| Laurent Pueyo | Space Telescope Science Institute |
| Yu-Jing Qin | California Institute of Technology |
| Garreth Ruane | Jet Propulsion Laboratory |
| Steph Sallum | University of California, Irvine |
| Julia Scharwaechter | Gemini Observatory, NOIRLab |
| Dirk Schmidt | National Solar Observatory |
| Garima Singh | Gemini Observatory, NOIRLab |
| Suresh Sivanandam | University of Toronto |
| Gaetano Sivo | Gemini Observatory, NOIRLab |
| Martin Still | National Science Foundation |
| Maaike Van Kooten | NRC / Herzberg A & A Research Center |
| Shelley Wright | University of California, San Diego |

# Table of Contents

# 1. Executive Summary

The United States Decadal Survey on Astronomy and Astrophysics, Pathways to Discovery in Astronomy & Astrophysics for the 2020s (Astro2020), identified the most compelling astrophysical science goals warranting national investment over 2023-2032 and beyond. Sponsored by the National Aeronautics and Space Administration (NASA), National Science Foundation (NSF), Department of Energy (DoE) Office of High-Energy Physics, and Air Force Office of Space Research (AFOSR), and solicited by the National Academies of Science, Astro2020 put forth a broad, integrated ground- and space-based vision for frontier advancement of astrophysical knowledge within three major themes: Worlds and Suns in Context, New Messengers and New Physics, and Cosmic Ecosystems (Figure 1, top-panel). Within each theme, Astro2020 recommended priority science areas, new programs and augmentations, and foundational developments.

Adaptive optics (AO) was explicitly cited in Astro2020 as an essential observational technique supporting multiple top-level priorities. The foremost ground-based priority was identified as "large (20 – 40 m) telescopes with diffraction-limited adaptive optics", while high-contrast AO was noted as an important technology risk reduction role in the development of the first space mission to enter the Great Observatories Mission and Technology Maturation Program (i.e. "a large (~6 m aperture) infrared/optical/ultraviolet (IR/O/UV) space telescope"). Astro2020 further identified AO as an exemplar of the importance of mid- and small-scale investments in ground facilities with the quote, "The development of adaptive optics (AO) in the 1990s serves as an excellent case in point, and led to ground-breaking advances such as the direct imaging of exoplanets, and the precise definition of the orbits of stars that determined the gravitational forces near the black hole at the center of the Milky Way." AO remains a young and vibrant field with much more to contribute to answering the pressing scientific questions of our time.

Rapidly evolving AO application areas and technologies compels the AO community to assess and communicate the science priority areas and new AO observational capabilities of greatest promise toward advancing Astro2020 goals. To this end, the US AO community held a workshop entitled, "*An Adaptive Optics Community Strategic Response to Astro2020*", May 16–18 2023. This report is the product of that workshop and is intended to identify future AO capabilities needed to most fully realize Astro2020 science goals. Sponsored by NSF's NOIRLab, the workshop included 28 in-person and 14 virtual representatives from the AO science and technology communities, public and private institutions, and numerous observatories including both US-ELTs. Representatives from the NSF/AST, DoE/LLNL, and NASA/GSFC research programs also attended, while in some cases also providing advance written input in response to canvassing for complementary mission needs.

We summarize the areas of greatest potential science impact within each Astro2020 theme, along with the correspondingly required AO systems and innovations, in Figure 1 (middle and bottom panels, respectively). Reinforcing key findings of Astro2020, we find that ground-based AO is vital to an integrated national astrophysics program, providing necessary precursor science and mission definition data and enabling risk mitigation activities in advance of the Astro2020 portfolio. During and after mission execution, AO plays an equally important role in providing unique and complementary science data at wavelengths and spatial resolutions critical for the full scientific understanding of targeted phenomena.

More specifically, we identified the following AO systems/capabilities as critical to fully realizing Astro2020 science goals, in order of priority:

1. Completion of the US-ELTs (GMT and TMT) with their first generation AO systems and science instruments (identified as Astro2020's top ground-based priority). US-ELT AO systems represent one of the largest single advances towards answering Astro2020's key scientific challenges.
2. The following AO capabilities:
    a. Primarily motivated by Worlds and Suns in Context, narrow-angle, high-resolution systems for exoplanet and high-contrast observations with 8-m class and ELTs, especially at visible wavelengths, with the goal of characterizing exoplanets with planet:star flux ratios of ~ $10^{-7}$ – $10^{-9}$ at angular separations of of < 3 λ/D with a target of 1 λ/D.
    b. Motivated by the New Messengers and New Physics, development and availability of a rapid response AO network on 2-m to 8-m class telescopes, at both enhanced seeing and diffraction-limited resolution, to follow-up on transient and dynamic sources discovered by Rubin, LIGO and similar systems.
    c. Motivated by both New Messengers and New Physics and Cosmic Ecosystems:
        i. Visible light AO systems for 8-m class telescopes with 0.015 arcsec imaging resolution or better across fields of view of 30 arcsec diameter or more.
        ii. Enhanced seeing systems for both single and multi-object spectroscopy and imaging. Wide-field systems will improve observing efficiency (2x or better) over fields-of-view of 6 arcmin diameter or larger.
    d. Motivated by all three science challenges:
        i. Improvements to quantitative science (e.g. astrometry, photometry) via better calibrations and PSF-reconstruction.

ii. More transparent and easier to use AO systems to enable efficient science by a broader user community (especially since the ELTs will be near 100% AO).
iii. 100% sky coverage which will require novel approaches to guide stars and wavefront sensing.

3. Development and maintenance of a sufficiently large and diverse AO Workforce (including PI Scientists, Support Scientists, Postdoctoral Researchers, Graduate Students, Optical, Hardware, and Software Engineers and Support Technicians. This Workforce should include academic and observatory staffing for system development and operation.

To achieve the above capabilities we recommend coordination, cooperation, and collaboration between the major US AO research groups as well as with the international community. Coordinated joint funding avenues should be investigated from public and private sources. The next generation of more capable AO systems will require technical advances in multiple areas (e.g. adaptive secondary mirrors, artificial beacons, wavefront sensors and control techniques) and the implementation of reliable, standardized AO components (e.g. lasers) where practical.

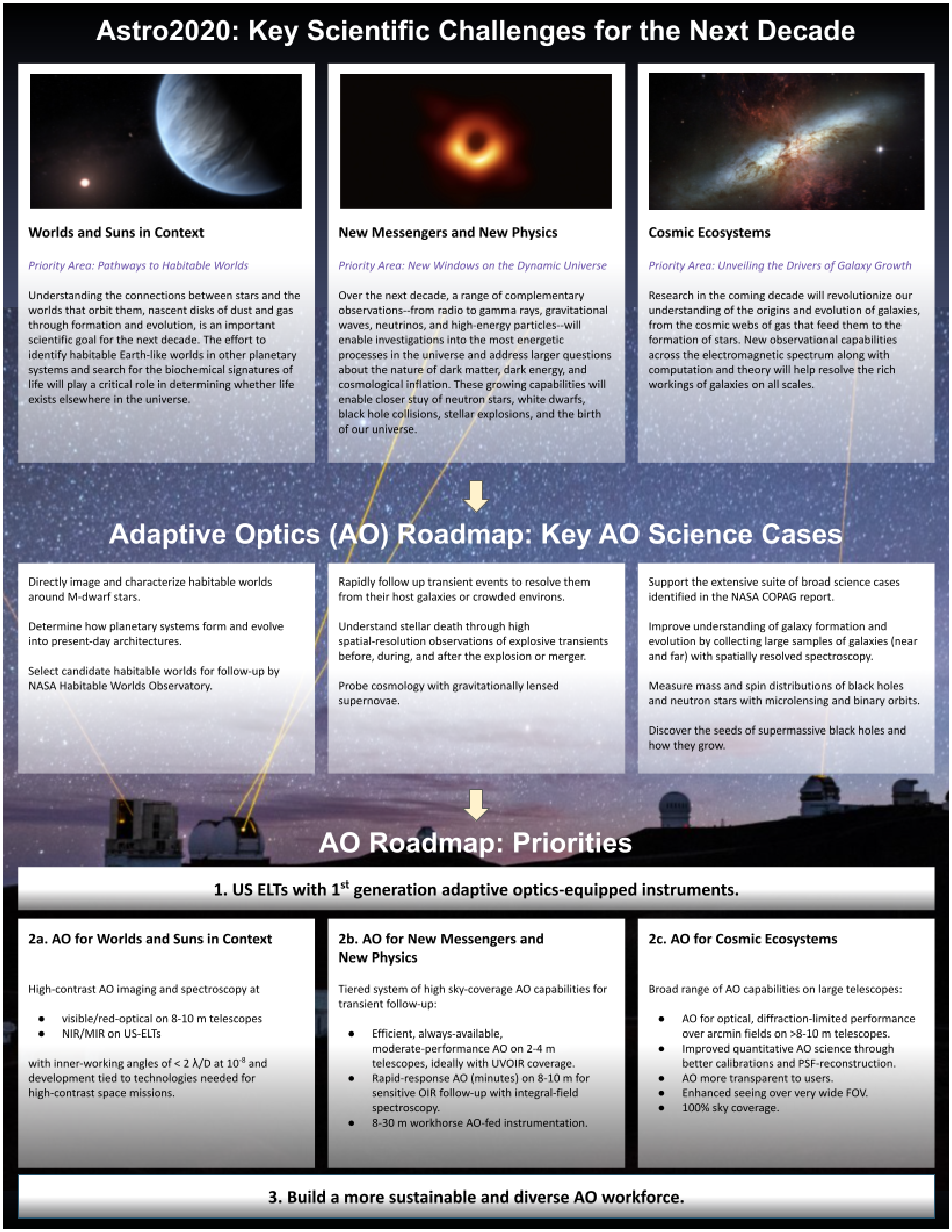

**Figure 1:** Top-panel: Astro2020 Key Science Cases (extracted from the Astro2020 Interactive Report). Middle-panel: A selection of key Astro2020 science cases where AO can play essential roles. Bottom-panel: Recommended AO priorities to achieve the Astro2020 Science.

## 2. Adaptive Optics and the Astro2020 Science Cases

Since the close of the 20th Century, Adaptive Optics (AO) has been demonstrated to be an extremely effective and valuable tool for ground-based astronomy and there are a variety of AO systems available to the global astronomical community on nearly every major US and non-US Optical/infrared (OIR) observatory. AO has benefited and enabled many areas of astrophysics, including Solar Physics, Solar System planetary science, dynamical and metallicity studies of low and mid-redshift galaxies, gravitationally lensed studies of dark matter and energy, exoplanets, stellar populations, as well as the discovery of and confirmation of the black hole in our own galaxy. As such it has produced a large number of high-impact science. The next generation of Extremely Large Telescopes (ELTs) have therefore been designed around AO systems to make use of high-spatial resolution in addition to their large collecting area.

In this section we present the three key scientific challenges and priority areas identified in Astro2020 and, as summarized in Figure 2, how ground-based AO observations will contribute to addressing these challenges.

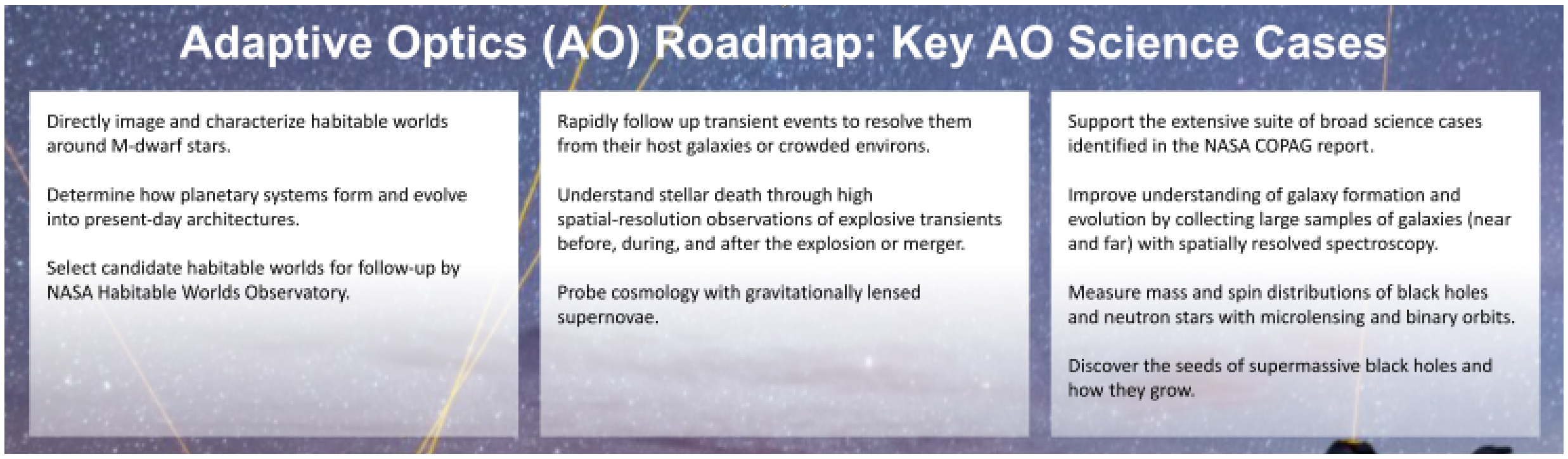

**Figure 2:** Key AO science cases flowed down from Astro2020.

### 2.1. Worlds and Suns in Context

Within the science theme of “Worlds and Suns in Context”, Astro2020 identified the Priority Area “Pathways to Habitable Worlds”. The ultimate goal of this area is to search for life elsewhere in the universe by identifying potentially habitable exoplanets, and assess whether their atmospheres contain the biochemical signatures of life. To meet these goals we must understand “the connections between stars and the worlds that orbit them, from nascent disks of dust and gas through formation and evolution”.

As one of the keys to achieving these goals, Astro2020 identified "ground-based extremely large telescopes equipped with high-resolution spectroscopy, high-performance adaptive optics, and high-contrast imaging and named the US-ELTs as the number one ground-based priority. The US-ELTs promise to revolutionize the study of star and planet formation, particularly through diffraction-limited and high-contrast imaging. The planned TMT (30 m) and GMT (25 m) offer AO-enabled diffraction-limited spatial resolutions four to five times better than existing ground-based telescopes and current and planned space telescopes. For background-limited observations, which includes high-contrast imaging through AO-corrected turbulence, the exposure time for equivalent observations scales as the fourth power of the telescope diameter ($D^4$), resulting in dramatic improvements in ground-based sensitivity with the ELTs.

For "Worlds and Suns in Context" we identified several key science cases for AO-equipped ELTs:

- Directly image and characterize habitable worlds around nearby M-dwarf stars in reflected light. Since they are cooler than Sun-like stars (FGK), the habitable zones (HZs) of M-dwarfs are closer in. While smaller projected separation makes observations more challenging, the planet:star flux ratio improves markedly: from $10^{-10}$ to $10^{-7}$. Thus the larger diameter of ELTs (compared to planned space observatories) will enable observations of planets in M dwarf HZs.

- Directly image and characterize habitable worlds around nearby Sun-like stars in thermal emission. At ~10$\mu$m the planet:star flux ratio is a much more favorable ~$10^{-7}$, and the speckle contrast scales as $1/\lambda^2$. Combined with the $\lambda$/D IWA and $D^4$ sensitivity improvements which will be provided by the ELTs, HCAO instruments on US-ELTs will enable observations of planets in FGK star HZs.

- Imaging of self-luminous giant planets and brown dwarfs and characterization at medium to high spectral resolution to determine how planetary systems form and evolve into present-day architectures. Such observations will extend work already started with existing telescopes, but use the improved resolution and sensitivity of the ELTs to fill in gaps in the exoplanet distribution. Such observations will be conducted from the near-IR through N band (~10 um), and allow studies of their formation history and system evolution. Observations of variability will enable studies of planetary rotation and cloud dynamics.

- Imaging of planet formation in situ: studies of protoplanetary disks, debris disks, and accreting planets. The dramatic improvement in IWA will enable studies of circumstellar material at currently inaccessible separations and the improved sensitivity of AO-equipped ELTs will improve observations of faint disk structure. The improved IWA and sensitivity will further studies of accreting protoplanets and the connections between them and the structures observed in protoplanetary disks.

The science goals mentioned above are aligned with those of the NASA Habitable Worlds Observatory (HWO), which is a ~6m diameter IR/O/UV telescope with high-contrast (10-10) imaging and spectroscopy. HWO was recommended by Astro2020 as NASA's next flagship Astrophysics mission. Representatives from the NASA Exoplanet Exploration Program Analysis Group (ExoPAG) were invited to the workshop. The workshop consensus was that there is strong synergy between ground-based and space-based high contrast imaging, and that technology development in each regime benefits the other directly. Observationally, ground-based high-contrast imaging will directly contribute to the goals of the HWO through the additional science case:

- Reconnaissance of HWO exoplanet imaging targets to assess whether the habitable zones are free of confusion, are dynamically stable. In addition to high-contrast coronagraphic imaging, diffraction-limited AO observations will contribute to a broad range of observations including astrometry and radial velocity measurements. Existing ExAO systems can be used for much of this precursor work, especially with the push to visible wavelengths and will help identify the final target list for HWO.

The consensus of the workshop was that, in order to achieve the above science goals the community should focus on the development of High Contrast AO (HCAO) which requires ExAO performance. Achieving these goals requires systems that achieve higher contrast ratios at smaller separations and work at both long and short wavelengths. Specifically, we recommend development of HCAO imaging and spectroscopy for:

- Visible to NIR HCAO on US-ELTs (ExAO). This will enable the detection and characterization of terrestrial planets orbiting nearby M-dwarf stars. This program should exploit existing instruments on current 8-10 m telescopes for developing and demonstrating technology for the US-ELT's.
- Thermal/MIR wavelength HCAO on US-ELT's. This has the potential to provide the first direct images of rocky exoplanets orbiting Sun-like stars.

The community would benefit from continued coordinated development of (HCAO) technologies for ground-based high-contrast imaging and spectroscopy, in particular to ensure that ExAO instruments are ready to capitalize on the US-ELTs. Examples of technologies in need of development include

- Deformable MIrrors: large format (up to 250x250 actuators) deformable mirrors are needed to provide the level of turbulence correction required for ExAO on the US-ELTs.

- Detectors for WFS: high speed (at least 2000 FPS), low noise (less than 1 electron per read), large format (512x512 pixels) detectors are needed to sense the wavefront at sufficient sampling to provide the level of turbulence correction required for ExAO on the US-ELTs.

- Algorithms: development of algorithms for wavefront reconstruction, control laws, and post-processing are needed. The rapid development of machine learning/AI and non-linear techniques shows great promise.

A more extensive list of technology development needs is given Appendix B.5.

HCAO systems can be defined as producing corrected PSFs with Strehl ratios of at least 80% on bright targets, which require high-order wavefront sensing and control coupled with a high update rate (generally 1 kHz or greater). HCAO systems are in regular operation on 6.5 to 8-m class telescopes. These instruments make use of coronagraphs for high-contrast imaging and spectroscopy similar to those planned for future space telescopes.

Over the last decade or so, there have been major developments in adaptive optics, coronagraphy, optical manufacturing, wavefront sensing, and data processing which have brought about a new generation of high-contrast imagers and spectrographs on large ground-based telescopes. The existing ExAO systems available to the US community include coronagraphic imaging and spectroscopic capabilities at Gemini, Keck, Subaru, LBT, & Magellan are summarized in Table E1.

For the “Pathways to Habitable Worlds” science cases we highlight, today’s systems are limited primarily by telescope diameter. ExAO instruments are in development for the TMT (MODHIS and the TMT-PSI concept) and the GMT (GMagAO-X, which is in the preliminary design phase). The construction of the US-ELTs is crucial to achieving these science goals, and the workshop consensus was that the US-ELTs should prioritize ExAO and high-contrast science.

## 2.2. New Messengers and New Physics

Within the science theme "New Messengers and New Physics", Astro2020 identified "New Windows on the Dynamic Universe" as the priority area focusing on the exploration of transient and explosive phenomena via space and ground-based facilities. This priority area builds on time-resolved detections across the electro-magnetic (EM) spectrum and combined with particle and gravitational wave signals to study the nature of stellar explosions, neutron star and black hole mergers, or tidal disruption events near supermassive black holes.

The start of Vera C. Rubin Observatory's operations (~2025) and the launch of the Nancy Grace Roman Space telescope (~2027) will further amplify the role of time-domain research in modern astronomy. Rubin Observatory's Legacy Survey of Space and Time (LSST) is expected to deliver up to 10 million transient alerts per night from seeing-limited multi-color imaging. Further classification of these alerts will require flexible follow-up options with a range of instrument modes across the EM spectrum.

While high-angular-resolution capabilities have traditionally been underutilized for time-domain and multi-messenger applications, they offer key advantages for characterizing transients against the contamination from their underlying host galaxy light or for identifying the location of transients in crowded stellar regions or near active galactic nuclei. High-angular resolution further enables studies of lensed transients that remain unresolved in seeing-limited observations or the search for transient precursor stars by comparing astrometrically calibrated pre- and post-event data.

Space telescopes offer less flexibility for time domain science because of pointing and slew time restrictions. Ground-based AO systems can provide a greater flexibility to achieve high angular resolution, especially for science cases requiring rapid response times (on the order of a few minutes) or high-cadence monitoring.

The consensus from the workshop was the recommendation of a tiered scheme of AO systems on telescopes of varying aperture to support the time domain and multi-messenger priorities of Astro2020 by 2030:

- 20m - 40m telescopes:

  Build the US-ELTs equipped with workhorse AO-fed instruments in the NIR with high sky coverage to follow up the faintest, most interesting transients too faint for smaller telescopes.

- 8m - 10m telescopes:

  Investment in high-cadence and rapid-response AO systems with as high a sky coverage as possible, which provide sensitive optical-to-IR follow-up capabilities with a focus on integral field spectrographs to support transient localization and contextualization. The sensitivity gains of advanced AO on these apertures *are the only way* to spectroscopically characterize and understand transient phenomena at the r ~ 25 single-visit brightness limit of LSST.

- 2m – 6m telescopes:

  Deployment of efficient, moderate-performance networks of AO systems - ideally working in the UVOIR range - with high sky coverage and continuous nightly availability. Such systems allow the exploration of various survey cadences, deep-drilling, and synoptic characterization of brighter targets, essential to placing the faintest transient discoveries into a scientific context.

The main technology and software priorities enabling this tiered approach are:

- AO systems with *high sky-coverage* to reach transients wherever they occur.
  - Such systems will require LGS; thus it is important to minimize overheads (both setup and real-time) due to LGS closures from aircraft and satellites
  - To maximize the solid angle of sky over which image quality improvement is available, efforts to increase the accuracy of atmospheric tip/tilt correction using optimized and novel sensing techniques deserve support.
- AO systems with *high nightly availability* and support for Target of Opportunity (ToO) interrupts.
  - Development of low-cost AO systems that can be deployed on smaller 2m to 4m class telescopes with large observing windows dedicated to time-domain science.
  - Advancement in robotic AO systems with the ultimate goal of routinely deploying fully autonomous systems on 2m and 4m class telescopes.

- AO systems that are *efficient* with high throughput.
    - Development of moderately-wide to wide field AO technology (LTAO, GLAO, MCAO, MOAO) to support quick searches of large sky areas.
    - Optimization of time-domain broker services for the specific use with AO.
- AO systems that are reliable and robust, *user-friendly* (turnkey) AO systems available for ToOs.
- AO systems with *broad wavelength coverage* across the optical and NIR. This will require research and development for visible AO.

To maximize time-domain synergies with other facilities, the ground-based AO development should prioritize telescopes that provide at least partial overlap with the regions of the sky accessible from Rubin Observatory, the suite of Astro2020 space telescopes, and/or other discovery facilities.

Several ongoing projects contribute toward, but will not fully satisfy, the time domain AO needs described above:

- The Keck All-sky Precision Adaptive-optics (KAPA) project with expected first-light in 2024 will use multiple laser guide stars to deliver narrow-field (~20”), diffraction-limited (FWHM~50 mas), infrared images with a sky coverage of >50% to the existing integral field spectrograph OSIRIS and planned Liger integral field spectrograph in 2027.
- The next-generation Gemini-North AO (GNAO) facility with expected first light in late 2027 will have similar capabilities to KAPA as well as a modest-field (up to 2’) enhanced seeing (FWHM ~ 0.12”-0.2” in K – J bands) mode. GNAO is specifically designed for time-domain applications with queue scheduling.
- Sharpening Images using Guidestars at the Hale Telescope (SIGHT) will provide LGS enhanced seeing (FWHM ~ 0.25”-0.4”) in visible through NIR bands on the 5-m telescope in California, commissioning no earlier than Fall 2025.
- The Robo-AO project has demonstrated robotic visible AO for 2-m class telescopes capable of surveying thousands of objects at high angular resolution. Robo-AO-2 (in construction in Hawai’i) and other similar facilities deliver essential, always-available capabilities for supporting time-domain science.

Additions to the national portfolio needed to provide complementary high-resolution imaging spectroscopy would include a publicly-available, autonomous, diffraction-limited AO survey system on a 2-4-m class telescope, routinely delivering better than 50 mas visible/NIR image quality. Similarly, an enhanced seeing survey system on a 8-m class telescope having access to the Rubinsky would multiply the nation's rapid LSST classification capability Finally, moderate performance visible-light diffraction-limited AO on 8-10m telescopes will be the only way to fully interrogate LSST transients at r ~ 25 or fainter using medium to high-resolution spectroscopy until the advent of the US-ELT's in the 2030's.

## 2.3. Cosmic Ecosystems

Astro2020 identified understanding "Cosmic Ecosystems" as a key challenge, noting,

*"The universe is characterized by an enormous range of physical scales and hierarchy in structure, from stars and planetary systems to galaxies and a cosmological web of complex filaments connecting them. A major advance in recent years has been the realization that the physical processes taking place on all scales are intimately interconnected, and that the universe and all its constituent systems are part of a constantly evolving ecosystem. … Unraveling the nature of this connection is one of the key science goals of the decade."*

"Unveiling the Drivers of Galaxy Growth" was identified as the Priority Area:

*"Research in the coming decade will revolutionize our understanding of the origins and evolution of galaxies, from cosmic webs of gas that feed them to the formation of stars. New observational capabilities across the electromagnetic spectrum along with computation and theory will help resolve the rich workings of galaxies on all scales."*

The NASA Cosmic Origins Program Analysis Group (COPAG), in active consultation with the Cosmic Origins community, is in the process of articulating Astro2020 science cases into lists of > 25 science gaps. The science cases cover a broad range of topics:

- Compact Objects and Energetic Phenomena
- Cosmology
- Galaxies
- Interstellar Medium and Star and Planet Formation
- Stars, the Sun, and Stellar Population

Although the specific science gaps are still being formulated, many of these can be addressed with sensitive, high spatial-resolution observations at OIR wavelengths. Broadly, the areas that fall within Cosmic Ecosystems and where AO observations play a key role are (1) understanding the death and afterlife of stars, (2) the formation and evolution of galaxies and their supermassive black holes, (3) the nature of dark matter and dark energy, and (4) stellar birth and the connection to planet formation. Additionally, the Planetary Science 2023 decadal review highlighted the need to understand the (5) dynamic atmospheres of Solar system planets and moons where ground-based AO will play a key role in time-domain studies and in supporting robotic space missions.

Within these broad science areas, we highlight three key cases where AO will play a significant role in the coming decade:

1. Improve understanding of galaxy formation and evolution using large samples of galaxies (near and far) with spatially resolved imaging and spectroscopy to measure stellar and gas morphologies, kinematics, and abundances.

   This requires AO systems that deliver a range of spatial resolutions from enhanced seeing ~0.3” resolution to optical, diffraction-limited AO on the ELTs with 0.015” resolution. Critically, the AO systems need to have high throughput to reach the faintest galaxies and high sky-coverage for out-of-the-plane, extragalactic survey fields.

2. Measure mass and spin distributions of black holes and neutron stars with gravitational microlensing and binary orbits.

   While gravitational wave facilities will deliver mass and spin measurements for large samples of black holes and neutron stars in distant galaxies, these results can only be placed in context using Milky Way samples of BHs and NSs found with astrometric gravitational lensing or binary orbits measured with radial velocities or astrometry. This requires AO systems with the highest possible spatial resolution for precise astrometry and to target crowded regions of the Galaxy where BHs, NSs, and microlensing events are most numerous.

3. Discover the seeds of supermassive black holes and how they grow.

   SMBH seeds may silently lurk in the centers of globular clusters or dwarf galaxies as present-day intermediate mass black holes (IMBH). Spatially resolved kinematics, either astrometry or radial velocities, are needed to find the dynamical signposts that point to a black hole’s location. This requires AO systems with the highest possible spatial resolution to (a) resolve globular clusters and the IMBH sphere of influence in galaxies in the local group or (b) deliver precise astrometry and radial velocities of individual stars in the centers of dwarf galaxies. The AO systems need to have high throughput and high sky coverage to reach faint targets well outside the Galactic plane.

Based on the common needs in the broad science areas and the three highlighted science cases, our top recommendation on the AO roadmap is

- The construction of the US ELTs equipped with their planned first-generation AO systems.

This will deliver high spatial resolution (FWHM < 20 mas) at infrared wavelengths. The Cosmic Ecosystems science cases are extensive and existing 8-10 m telescopes can serve to address many of these areas if equipped with the following AO capabilities:

- Visible-light, diffraction-limited AO that provides comparable spatial resolution (<20 mas) and fields of view (1-2 arcmin) to the ELT's IR AO systems.
- High throughput, wide field (>0.1 deg) AO with enhanced-seeing spatial resolution (~0.2") to improve survey sensitivity.
- High sky-coverage for the above AO systems with a target of 100% through the development of
    - More powerful artificial laser guide stars
    - Orbital guide stars
    - More sensitive visible and IR tip-tilt sensors

- Improved quantitative science from AO observations (e.g. astrometry, photometry, morphology) with better calibration and telemetry systems and PSF reconstruction.

Ground-based AO systems will complement OIR observations from JWST, Rubin, Roman, and Euclid (see Appendix F and G for details). In particular, ground-based AO systems will excel at providing the highest spatial resolutions and feeding a wider variety of instruments when compared with these complimentary facilities.

# 3. AO Development for Astro2020 Science Goals

The science cases outlined in Section 2 motivate the implementation of several planned and new AO capabilities. As summarized in Fig. 3, the top-level recommendations that flow from the science cases include:

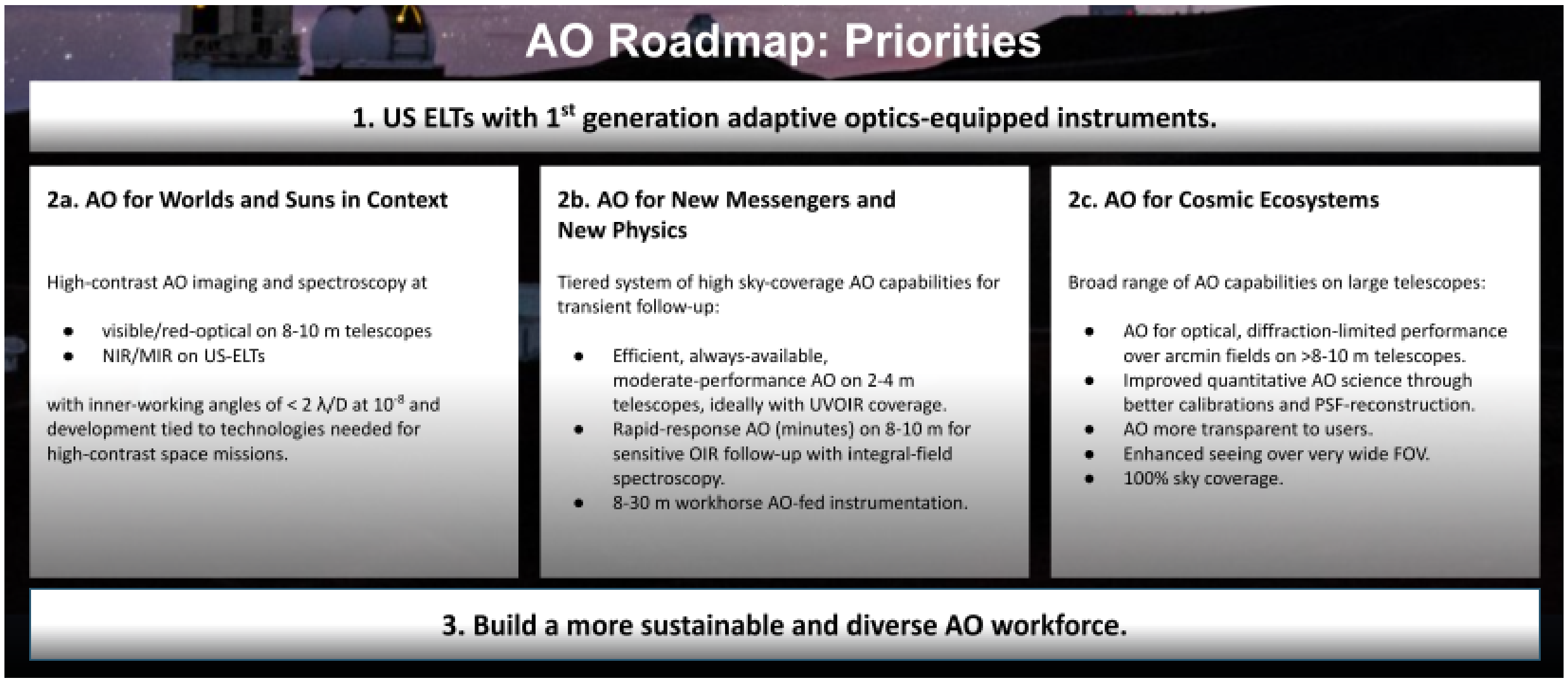

**Figure 3:** AO roadmap priorities to achieve the Astro2020 science challenges.

In the following sections, we present additional detail on the technical capabilities needed to achieve each of these recommended AO systems and the overall Astro2020 objectives. For context, a description of current AO facilities and capabilities is presented in Appendix D and a more detailed discussion of the future directions for AO Technology and Operations is presented in Appendix B.

## 3.1. US ELTs with First-Generation AO Systems

The AO capability that will have the largest impact on Astro2020's key scientific challenges is the completion of the US-ELTs (GMT and TMT) with their first generation AO systems and science instruments (identified as Astro2020's top ground-based priority). The telescopes and first-generation AO systems are in late stages of design and technologies are in place. However, additional studies of optimal operation, calibration, and science extraction methods performed on smaller telescopes can have an impact on the ELT's AO systems (see Section 3.6).

The currently planned first-generation AO systems on the US ELTs do not include high-contrast systems designed to detect and characterize terrestrial planets orbiting M dwarfs. For the Pathway to Habitable Worlds science case, this is a key limitation that we address in Section 3.2.

## 3.2. High-Contrast Visible to MIR AO on US ELTs

Primarily motivated by Worlds and Suns in Context, we identify the need for high-contrast imaging and spectroscopy with AO systems providing narrow-FOV, high-resolution correction on bright natural guide stars with an ELT class telescope from visible to mid-infrared wavelengths. Such systems would ideally reach contrasts of ~ $10^{-7}$ – $10^{-9}$ and inner working angles of < 3 λ/D with a target of 1 λ/D.

On the larger aperture of ELTs, instruments with such specifications will enable detection and characterization of terrestrial, potentially habitable, exoplanets orbiting nearby stars. To reach this goal, we must leverage technology development with current HCAO facilities. AO instrument concepts for the ELTs, including MODHIS, PSI, GMagaO-X, and TIGER have the potential to provide high impact observations of rocky planets around the nearest stars. We recommend the continued development of HCAO systems for the US-ELTs that support observations from the MIR, NIR, and into the visible as the technology is proven.

The development of future ExAO high-contrast systems for ELTs introduces a number of technological challenges in terms of instrumental hardware and algorithms and software for both control and analysis. The combination of high-order wavefront sensing and control, coronagraphy, post-coronagraph wavefront control, and post-processing must be able to characterize planets with planet-to-star contrast ratios of ~ $10^{-7}$ – $10^{-9}$ to achieve the goals of the Decadal Survey.

1. Deformable Mirrors:
   For the ELTs, the most pressing need is the development of large actuator count deformable mirrors (DMs) required to achieve > 80% Strehl ratios at short wavelengths. DMs with ~250 x 250 actuators, corresponding to a 12 cm spacing on a 30 m aperture, are needed. Such a DM will require a peak-to-valley stroke of up to ~ 5μm surface running at a minimum of 2kHz with a final goal of 5kHz –10kHz.
2. Detectors for Wavefront Sensing:
   To measure the wavefront at these scales, large format, high-speed, low-noise detectors will be required, corresponding to at least 512 x 512 pixels sampling at a minimum rate of 2 kHz, with a goal of 5 kHz to 10 kHz, and read noise less than $1e^-$. Low latency and a 100% duty cycle are required and the use of a global shutter should be investigated and may also be required.
3. Wavefront Sensors:
   Further development of new wavefront sensing techniques, and optical design and fabrication improvements, are needed to maximize both sensitivity and dynamic range. Photonics-based sensors show exciting promise and need to be proven in on-sky use.

4. Wavefront Reconstruction:
   Wavefront reconstruction algorithms, including especially nonlinear algorithms, need to be further developed to improve the dynamic range of the most sensitive (photon efficient) wavefront sensors. Advances in machine-learning algorithms show great promise in this area.
5. Control:
   Optimized control laws, including prediction and perhaps nonlinear algorithms, need to be further developed and demonstrated in routine use on-sky. The use of sensor fusion (e.g., accelerometers, on-instrument wavefront sensing) as part of real-time wavefront control should be investigated.
6. Coronagraphs:
   Coronagraph technology needs to be improved to routinely obtain inner working angles (IWAs) of < 3 λ/D with a target of 1 λ/D, and must be demonstrated on current instruments. Continued development of coronagraph optical design and manufacturing is needed to achieve this. Photonics-based coronagraphs offer the possibility of sub-λ/D IWA, and should be demonstrated on-sky.
7. Wavefront Control:
   The ability to suppress quasi-static speckles with post-coronagraphic wavefront sensing and correction must be developed for routine on-sky use by current instruments. Daytime and on-sky calibration techniques should be developed and optimized to correct for non-common path aberrations.
8. Post-Processing:
   The use of AO telemetry, similar to conventional PSF reconstruction, offers great promise towards minimizing and eventually eliminating speckle noise. Such algorithms need to be developed and tested using on-sky data from current instruments.
9. Segment/Island Piston:
   ExAO systems are particularly susceptible to segment or island phase discontinuities, also sometimes called petal modes. This is already recognized in the "low-wind effect" on some current ExAO systems, and will be an important issue on the segmented GMT (due to its primary and secondary mirror segment design) and TMT (due to the primary mirror segments and secondary supports). Sensing and control strategies for segment piston discontinuities must be developed and demonstrated on-sky.

The workshop consensus was that HCAO/ExAO instruments on existing large telescopes (such as SCExAO, MagAO-X, KPIC, LBTI, and GPI 2.0) should continue to be used for the development and on-sky demonstration of these technologies.

## 3.3. Efficient, All-Sky AO on Small-to-Mid Sized Telescopes for Time Domain Astronomy

Motivated by the New Messengers and New Physics, we recommend the development and availability of a rapid response AO network on 2-m to 8-m class telescopes, at both enhanced seeing and diffraction-limited resolutions, to follow-up on transient and dynamic sources discovered by Rubin, LIGO and similar systems. The primary features of time domain astrophysics (TDA) AO systems are:

- *High sky coverage:* Transients occur throughout the sky at unpredictable times and locations. Thus TDA AO systems need to be equipped with laser guide stars (LGS) to address more than 50% and ideally 100% of the sky. In addition to the LGS, the AO system should support a wide field of regard for using any natural guide stars (NGS) needed for either tip-tilt or tip-tilt plus focus, which Rayleigh LGS and Sodium LGS don't provide respectively.
- *Always available:* Small to mid-sized facilities often have more flexible scheduling options, including ToO modes. A TDA AO system should support these modes. Optical LGS propagation may impact aircraft and require monitoring and frequent closures in regions of high air traffic; thus UV LGS are strongly preferred. Novel methods for obtaining all-sky US Space Force clearance for LGS propagation have been successful for Robo-AO and should be implemented more widely.
- *Sensitive:* The advantages of a TDA AO system is to moderately improve spatial resolution for resolving transients from their galaxy environments *and* to improve the sensitivity and efficiency by reducing sky background. This favors AO systems implemented with deformable mirrors built into the telescope (e.g. deformable secondary) and without additional optical relays in the science path.
- *Low-cost:* The diverse needs of TDA follow-up favor more AO systems deployed on more telescopes with possibly more flexible performance requirements. This can only be achieved if the cost of an AO system drops. Cheaper technologies exist, but have not yet been fully realized in an AO system, and include CMOS-based wavefront sensors, UV lasers, and MEMS deformable mirrors. The cost of large-format DMs to serve as adaptive secondary mirrors would need to drop to achieve an ideal TDA AO system.

## 3.4. Visible-Light, Diffraction-Limited AO on 8-10 m Telescopes

Motivated by both New Messengers and New Physics and Cosmic Ecosystems, we recommend a visible light AO system for 8-m class telescopes with <0.02 arcsec imaging resolution or better across fields of view of 30 arcsec diameter or more. Such a system should have high sky-coverage (>50%) to reach extra-galactic targets and transients wherever they might occur. To reach near-diffraction-limited performance on 8-10 m telescopes, there are several key technology advances needed and described below.

First, *higher actuator count deformable mirrors* with >2000 actuators are necessary and can already be purchased either as a large-format (>0.5 m) adaptive secondary mirror, medium-format (0.1-0.25 m) flat mirrors for AO systems at Cassegrain or Nasmyth locations, and small-format (<5 cm) flat mirrors with less stroke for compact, narrow-field, or second-pass AO systems. To reach a >30 arcsec field of view, multiple deformable mirrors conjugate to atmospheric layers at different heights are needed. This is known as a multi-conjugate AO system (MCAO) and MCAO systems are currently deployed at Gemini (for NIR imaging) and the Inouye solar telescope.

Second, multiple *bright laser guide stars* are critical to achieve an AO system with visible capability, a >30 arcsec field, and high sky coverage. The LGS brightness must be high enough to provide sufficient flux to measure atmospheric aberrations at high spatial frequencies and fast AO loop times. Multiple beacons, ideally side launched from the telescope are needed to provide tomographic (3D) mapping of the atmospheric turbulence. Bright sodium lasers exist (e.g. TOPTICA 60W) to produce guide stars at 90 km; but, they are expensive. Exploration of cheaper, but still bright, Rayleigh lasers pulsed to generate guide stars at 15-20 km above a sight may lead to hybrid sodium + Rayleigh solutions that are significantly cheaper. One innovative solution to artificial guide stars are orbiting guide stars that can deliver very bright laser sources shining from a satellite that look identical to a natural guide star. Orbital guide stars have strengths and limitations that need to be fully explored through on-orbit testing.

## 3.5. Sensitive, Wide-Field, Seeing-Enhanced AO

Enhanced seeing systems for multi-object spectroscopy and imaging in the visible and NIR. These will both improve observing efficiency and productivity (2x or better) over fields-of-view of several arcminute diameter or larger. The primary features of these seeing-enhanced AO systems are:

- *High sky coverage*: Ability to observe large continuous portions of the sky.
- *Wide-fields of view*: Leverage seeing conditions at the sites for corrected fields of view of several arcminutes in diameter or larger.
- *Sensitive and available to all instruments*: High throughput and low emissivity with the ability to provide enhanced seeing to all science instruments on the telescope. This favors AO systems with the deformable mirror integrated with the telescope optics (e.g. deformable secondary mirror).

## 3.6. Improvements to Current AO Systems

Current AO systems are heavily used and would benefit from improvements to improve the precision and accuracy of the scientific measurements including photometry and astrometry. This can be achieved through better and more frequent optical calibrations of the AO system and PSF-reconstruction. Typically AO calibrations are performed once a day at most. However, access to extensive AO and telescope telemetry and advances in machine learning may provide better, real-time calibrations updated for changing conditions. PSF-reconstruction has been a holy grail for adaptive optics for many decades; however, several European teams have made significant progress and now routinely deliver reconstructed PSFs for each AO science image for a subset of VLT AO systems. Such methods should be implemented and improved for US AO systems as well.

Current AO systems are typically expensive to operate and maintain and are not always as robust and reliable as seeing-limited instruments. AO is also a complex observing mode that is not yet turn-key. Reducing costs, automating operations, and improving maintainability are essential to broaden the use of AO, especially for deployment on larger numbers of small- and mid-sized telescopes. Cost reductions may come from utilizing commercial equipment (e.g. CMOS detectors, UV lasers) that may have performance reductions of ~25%; but cost reductions of >10x. Automated operations often require a significant capital cost to upgrade old hardware and software, particularly on smaller telescopes. Advances in sensor fusion and artificial intelligence are needed to keep the AO system at peak optimization in real time without intervention. These advances can also be deployed to reduce the human-in-the-loop aspects of routine calibration and AO system maintenance. Developments in consumer AO and real-time image processing show that even complex systems can be made “transparent” to the user.

Lastly, sky coverage is often a major limitation for AO systems and research and development is required to find novel and cheaper ways to improve sky coverage. Example areas to investigate include using orbital guide stars, utilizing hybrid LGS systems with expensive sodium lasers at high altitude alongside cheaper Rayleigh lasers, exploring dual-color laser options for measuring tip-tilt and focus terms from the LGS directly rather than from a natural guide star, etc. Most of these will require on-sky tests to investigate their utility.

## 3.7. A More Sustainable and Diverse AO Workforce

Development and maintenance of a sufficiently large and diverse sustainable AO Workforce (including PI Scientists, Support Scientists, Postdoctoral Researchers, Graduate Students, Optical, Hardware, and Software Engineers and Support

Technicians. This Workforce should include academic and observatory staffing for system development and operation. This will require more stable and fewer “soft money” positions especially within academia to maintain institutional knowledge especially at major instrument building and design facilities. Workforce retention can be addressed by (1) longer postdoctoral researcher positions, e.g. 5 years instead of 3 years including a US ELT AO Fellowship program, (2) academic institutes creating and supporting longer-term and permanent technical support positions, and (3) alternative career paths to academia, such as at national research laboratories and related industries, or other applications of AO.

## 4.0 Preliminary AO Roadmap

AO can and should play multiple critical roles in achieving the science goals and top space and ground-based priorities of the Astro2020 report. Figure 4 below lays out a timeline for this report's recommended AO milestones in support of the Astro2020 goals and beyond. Appendix B.8 and Table B.1 includes a more detailed roadmap including development goals for component technologies. Consideration was given to milestones that should be demonstrated to inform the 2030 and 2040 decadal reviews.

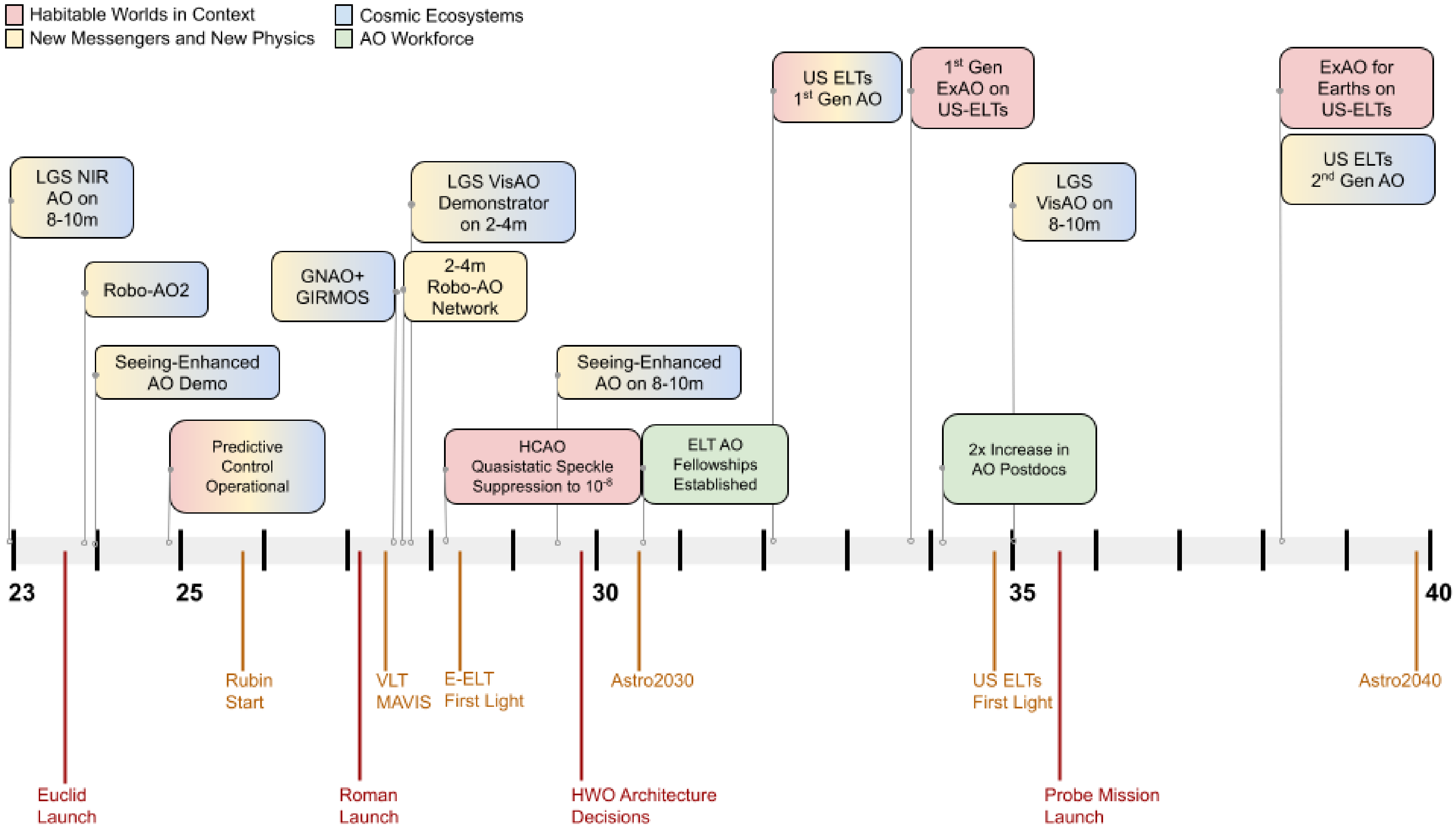

**Figure 4:** AO Roadmap Notional Timeline

# Appendices

The Appendices contain information in support of the main document. They are as follows:

A. COPAG Science Cases

   This presents an overview in the form of a table of the science questions to be addressed by Astro2020 and identifies how AO observations can contribute,  This relates to the Science cases described in §2. This information was presented by the COPAG representative at the Workshop.

B. Adaptive Optics Technology and Operations for the Astro2020 Science Goals

   This presents a detailed overview of the AO systems necessary to meet the Astro2020 science goals. It also addresses more general issues of AO operations and community policy. A detailed Preliminary AO Roadmap id also presented.

The following Appendices addresses the background of AO in astronomy and the US community in particular. The different types of AO systems are described as well.

C. Astronomical Adaptive Optics
   - C.1. Background and History
   - C.2. 2008 US AO Roadmap
   - C.3. Adaptive Optics Performance
   - C.4. Adaptive Optics Systems

D. Adaptive Optics Telescopes available to the US Community

E. The Current Adaptive Optics Landscape

F. Complementary Telescopes

G. Space Telescopes

## A. COPAG Science Cases

This presents an overview in the form of a table of the science questions to be addressed by Astro2020 and identifies how AO observations can contribute, This relates to the Science cases described in §2. This information was presented by the COPAG representative at the Workshop.

| Science Topics | Science Questions | Detailed Science Questions | Supporting Observations |
|---|---|---|---|
| **Compact Objects and Energetic Phenomena** | What are the mass and spin distributions of neutron stars and stellar mass black holes? | What do the mass and spin distributions tell us about neutron star and black hole formation? | AO observations, at visible and infrared wavelengths, of micro-lensing, Galactic stellar binaries, and identification of EM counterparts to gravitational wave sources. |
| | | What is the population of noninteracting or isolated neutron stars and stellar-mass black holes? | |
| | | What is the equation of state of ultra-dense matter? | |
| | Why do some compact objects eject material in nearly lightspeed jets and what is that material made of? | How do jets launch and accelerate? | AO systems can provide high-Contrast visible and IR observations of stars with jets (protostars), a rapid response to neutrino/GW events, and characterization of the stellar populations in the source environments. |
| | | What are jets composed of and how are particles accelerated within them? | |
| | | Are TeV and PeV Neutrinos and ultra-high energy cosmic rays produced in relativistic jets? | |
| | What seeds supermassive black holes and how do they grow? | How are the seeds of supermassive black holes formed? | High-spatial resolution AO observations of IMBH in nearby galaxies equivalent to synthesis radio imaging resolution. Identifying sources and internal motions in galactic centers and satellite galaxies. |
| | | How do central black holes grow? | |

| Science Topics | Science Questions | Detailed Science Questions | Supporting Observations |
|---|---|---|---|
| Cosmology | What are the properties of dark matter and the dark sector? | Dark sector signatures in small-scale structure | Visible and IR AO Observations of Dwarf Galaxies and Strong Lensing sources. Nucleosynthesis constraints from coupled high-spectral resolution observations of stars. |
| | | Dark sector imprints on Big Bang nucleosynthesis and recombination | |
| | What physics drives the cosmic expansion and the large-scale evolution of the universe? | End-to-End tests of Cosmology | Visible and IR AO observations looking at time-delay effects in gravitational lensing observing quasars and observations of detached eclipsing binaries for distance estimates. |
| | How will measurements of gravitational waves reshape our cosmological view? | Standard sirens as a new probe of the cosmic distance scale. | Rapid, high-spatial resolution visible and IR follow-up observations of transient events to identify and characterize EM counterparts of sources. |

| Science Topics | Science Questions | Detailed Science Questions | Supporting Observations |
|---|---|---|---|
| Galaxies | How did the intergalactic medium and the first sources of radiation evolve from cosmic dawn through the epoch of reionization? | Detailed thermal history of the IGM and the topology of reionization. | Visible and IR AO high-spatial resolution spectroscopy to determine the properties of nearby metal-poor populations. |
| | | Production of ionizing photons and their escape into the intergalactic medium | |
| | | Properties of the first stars, galaxies, and black holes | |
| | How do gas, meta.flow into, through, and out of galaxies? | Acquisition of gas necessary to fuel star formation | Visible and IR AO high-spatial resolution IFU observations of the IGM using quasar sightlines to characterize bulk properties of the gas flow on different scales. |
| | | The production, distribution, and cycling of metals | |
| | | The coupling of small-scale energetic feedback processes to the larger gaseous reservoir | |
| | | The physical conditions of the CGM | |
| | How do supermassive black holes form and how is their growth coupled to the evolution of their host galaxies? | The seeds of supermassive black holes | High-spatial resolution (AO) observations in the Visible and IR of the cores of external galaxies, with IFUs when possible. |
| | | The existence and formation of intermediate black holes | |
| | | Comprehensive census of supermassive black hole grown | |
| | | The physics of black hole feedback | |
| | How do the histories of galaxies and their dark matter halos shape their observable properties? | The dynamical and chemical history of the milky way | Visible and IR high-spatial resolution AO observations of resolved stellar populations of nearby galaxies to obtain the star formation history, coupled with moderate resolution (~1 km/s) spectroscopy of nearby galaxies to derive halo masses, and high spatial resolution IFU maps of galaxies at a range of z's -- moderate resolution to get spatially resolved chemistry. |
| | | The threshold of Galaxy Formation | |
| | | Connecting local galaxies to high redshift galaxies | |
| | | The evolution of morphologies, gas content, kinematics, and chemical properties of galaxies | |

| Science Topics | Science Questions | Detailed Science Questions | Supporting Observations |
|---|---|---|---|
| Interstellar Medium and Star and Planet Formation | How do star-forming structures arise from, and interact with, the diffuse ISM? | What sets the density, temperature, and magnetic structure of the diffuse ISM enabling the formation of molecular clouds; | Visible and IR high-spatial resolution AO observations with moderate-to-high spectral resolution with high SNR to identify diffuse interstellar bands (IR) or auroral emission (Vis). |
| | | How do molecular clouds form from and interact with their environment? | |
| | | How does injection of energy, momentum, and metals from stars (stellar feedback) drive the circulation of matter between phases of the ISM and IGM | |
| | How does gas flow from parsec scales down to protostars and disks? | How do dense molecular cloud cores collapse to form protostars and their disks? | High-spatial resolution IR AO observations of disks either in narrow band imaging (emission lines) where possible or with an IFU. |
| | | How do protostars accrete from envelopes and disks, and what does this imply for protoplanetary disk transport and structure? | |
| | | Is the Stellar Initial Mass function universal? | |
| | Is planet formation fast or slow? | What are the origins and demographics of disk substructures? | High-spatial resolution IR AO observations of disks either in narrow band imaging (emission lines) where possible or with an IFU. |
| | | What is the range of physical environments available for planet formation? | |
| | | How do turbulence and winds influence the evolution of structure in disks? | |

| Science Topics | Science Questions | Detailed Science Questions | Supporting Observations |
|---|---|---|---|
| Stars, the Sun, and Stellar Populations | What are the most extreme stars and stellar populations? | | Visible and IR high-spatial resolution AO imaging/spectroscopy of nearby stellar populations and embedded stellar populations |
| | How does multiplicity affect the way a star lives and dies? | | Visible and IR AO observations of binary systems across stellar evolutionary states and also identifying binaries with interesting stellar combinations. |
| | What would stars look like if we view them like we do the Sun? | | This requires interferometric resolution, e.g. CHARA, of nearby stars. Interferometric systems will require AO from the sub-apertures. Radial velocity measurements will characterize astero-seismic targets |
| | How do the Sun and other stars create space weather? | | Interferometric imaging of the surfaces of nearby stars and high contrast AO imaging of their environments. |

## B. Adaptive Optics Technology and Operations for the Astro2020 Science Goals

US AO developments over the last 15 years have been influenced and motivated by the last AO roadmap,*"A Roadmap for the Development of United States Astronomical Adaptive Optics"* published in April 2008, which was sponsored by the NSF's Division of Astronomical Sciences in response to the AO Community's request for a reliable public funding source for AO system research, development, and implementation.

Astro2020 identified the completion of the existing US-ELT first generation DL systems as the highest ground-based priority. It's important to note that these ELT AO systems have been defined for over 15 years and although advanced and groundbreaking, will represent older technology when the two telescopes become operational. What will be the second generation AO instruments for the US-ELTs? Given the rapid advances in AO over the last 15 years, development of the necessary technologies can be initiated on smaller aperture facilities with more ready access for development time which will assist in identification of the next generation ELT AO systems potentially in time for the 2030 Decadal Review.

The AO developments required to reach the Astro2020 science goals will be challenging. The following presents a detailed overview of the AO systems necessary to meet the Astro2020 science goals. It also addresses more general issues of AO operations and community policy.

### B.1. AO Facilities

Not all existing US AO-enabled telescopes are available to be used as developmental platforms. For example, the two Gemini telescopes are workhorse science instruments serving a highly competitive US and international community. Development work is ongoing at the predominantly private telescopes, e.g. Keck, LBT, Magellan, and is likely to continue at these sites. However, these telescopes are also scientifically competitive and pathfinder work, such as visible AO systems, could be developed on 4-m class telescopes for eventual upgrades to 8-m class facilities as discussed in §4.2.

One challenge facing the US AO community is that funding for future AO systems will be highly competitive given their cost, one example, the GNAO/GIRMOS will cost ~ $50M. Thus, development of multiple similar AO systems on the 8-m class telescopes is likely unrealistic, especially for NSF funding. Discussions at the workshop included whether the US science community could be better served by having specific AO capabilities on different telescopes or a dedicated AO telescope similar to the VLT's UT4. While we are able to provide technical responses, we are not able to make such strategic decisions for the US community.

## B.2. Visible Light Adaptive Optics

Astro2020 identified the development of Visible Light AO (VisAO) systems as being a critical growth area. VisAO will require both higher density deformable mirrors and an increased number of wavefront sensor sub-apertures versus existing systems. The increased number of sub-apertures will require brighter guide stars, coupled with low-noise, fast readout, high cadence detectors. It is important to note that a system built for diffraction-limited correction at 650 nm will only provide partial correction at the shorter blue and green wavelengths which will have an impact on the science so that a staged approach beginning with longer wavelengths is recommended.

There are different VisAO systems as there are existing NIR AO systems (see §A1.1.3). These include:

1. NGS High Contrast Systems
   Narrow-field, high-order correction with a goal of reaching a coronagraphic performance with an IWA of 3 λ/D and a contrast of $10^{-8}$.

2. LGS VisAO Systems
   This includes narrow-field (~ 10”) LTAO and moderate-sized field (~ 1’ – 2’) MCAO diffraction limited systems, as well as wide field (~ 10’) enhanced seeing GLAO systems. Both the LTAO and MCAO systems will provide diffraction-limited performance, except at the shorter wavelengths as mentioned above, and GLAO systems will provide enhanced seeing capabilities.

NGS VisAO systems will have a larger impact for ExAO and will require the same AO hardware improvements with DMs, WFSs and detectors as the LGS systems (see §B.5). In general, DM technology development would also include Adaptive Secondary Mirrors (ASMs), as well as standard optical bench DM upgrades. ExAO systems would benefit from both when an ASM is available, e.g. the planned use of ScExAO at Subaru. LTAO systems would also benefit from the development of both DM types especially for MOAO systems, e.g. GIRMOS/GNAO at Gemini N. In addition to the hardware improvements, there are also corresponding improvements required for AO control systems (RTC) to increase the speed and number of channels especially for tomographic reconstructions as used for LTAO and MCAO. High sky coverage is limited by NGS availability for tip-tilt and low order correction and there are a number of strategies which have been discussed to mitigate this issue and should be followed up with for the long-term goal of total sky coverage.

VisAO systems can be initially developed for 2-4m class apertures. The DKIST 4m Solar Telescope already has a successful operational VisAO MCAO system and the AFRL 3.5 m telescopes, at the Starfire Optical Range and the Maui Space Surveillance System have successful SCAO systems. Lessons learned from these systems for narrow- and wide-field applications can be used to develop 2-4m VisAO systems, depending upon applications, as pathfinders for an 8-m class system and which would complement the 8-m class NIR AO systems.

### B.3. Laser Guide Star Systems

The evolution of AO to shorter wavelengths and larger apertures will require more powerful sodium lasers. ESO is currently working with Toptica on the development of ~ 70W lasers compared with the existing 22W ones. There is also the option of a space-based source such as the NASA-Goddard ORCAS project with which Keck is partnering. It is recommended that research and development of both avenues of providing brighter artificial guide stars are continued and that the US community is engaged with the Toptica development and testing.

There are operational issues with ground-based lasers which affect observing efficiency. Laser propagation is subject to the following external events:

1. Satellite Laser Deconfliction

   This is an NSF requirement for US observatories and follows the US DoD Laser Clearinghouse (LCH) protocol of protecting all potentially vulnerable satellites from inadvertent illumination. The recent launches, since 2019, of LEO Satellite Constellations such as Starlink, have significantly increased the number of satellites and therefore the number of satellite closures. Although Starlink have recently waived LCH protection, over the next decade the number of LEO satellites is expected to increase significantly based on FCC filings ~ 100,000. With recovery times from a closure window ranging from 30s – 120s, depending upon the AO system, the number of LCH closures will cause a dramatic decrease in science time with LGS AO. Currently the US LGS propagating community are working with NOIRLab, the NSF, and the DoD to mitigate these closure windows with the goal of reclassifying the astronomical lasers to a lower impact category. This is also important for targets of opportunity (ToO) as the LCH currently requires target lists to be submitted a few days prior to observing in order to determine clearance windows. The use of alternate LCH approaches such as the partitioned sky regions, as opposed to target lists, used by Robo-AO is also recommended for ToOs and immediate follow-up to transient events as described §2.2.

2. Aircraft Protection

   Aircraft protection from inadvertent laser illumination is coordinated with the relevant civilian aviation administrations, i.e. the FAA (USA) and DGAC (Chile). Currently the use of Transponder Based Aircraft Detection (TBAD) has removed the requirement for human spotters and has automated the process of detecting a probable “collision” and shuttering the laser. However, unlike satellites, the commercial air traffic can vary considerably from site-to-site. For example, at Maunakea there are very few overflights per month (< 5) whereas US continental sites, such as Mt. Graham (LBT) and Mt. Hamilton (Lick), are close to the flight paths for nearby airports. Due to the large number of FAA closures, these observatories tend to not schedule LGS during the first half of the night. Chilean sites, such as ESO and Gemini S., have noticed increased air traffic over the last decade seriously impacting their propagation windows.

   The 355 nm UV Raleigh beacons, such as those used at SOAR and by Robo-AO, are aircraft safe, Class 1, as the maximum exposure time of an aircraft to the pulses reduces the Maximum Permissible Exposure (MPE) to ~0.1% – ~0.6%. . Given that, it is recommended that high-air traffic sites could consider using Rayleigh beacons, when the science goals permit, as an alternative to potentially increase their on-sky time. However, the trade-off is that the Rayleigh beacons produce a larger focal-anisoplanatism and increased wavefront error (reduced performance) than the sodium beacons which would significantly impact some science goals. It would serve to optimize the laser-propagating observatories to their environment as much as possible.

3. Laser Traffic Control

   Laser Traffic Control (LTC) exists at sites with multiple telescopes to prevent contamination of science data or the WFS on other telescopes. As an example, Maunakea favors a first-on-target approach, preventing the other telescopes from propagating if the “beams” intersect. Queue scheduling of telescopes and improved coordination between observatories would serve to minimize this impact.

4. Atmospheric Conditions

   Laser propagation generally requires photometric conditions so LGS AO systems will be more useful at more photometric sites. The telescopes should be able to quickly switch to LGS mode to take advantage of the changing conditions. This will require flexible scheduling in addition to queue scheduling from the telescope operations perspective.

### B.4. Wavefront Correction

#### B.4.1. Adaptive Secondary Mirrors

The use of an ASM as the wavefront corrector reduces the number of optical surfaces which increases optical transmission and reduces emissivity. It also provides access to a wider corrected field-of-view especially for GLAO systems. Routine ASM operational AO systems are currently available at the LBT and at one of the ESO VLT's, UT4. GMT and Subaru are in the process of procuring ASMs while Keck, Gemini and TMT are considering ASMs. GLAO systems, using ASMs, can increase science productivity at all telescopes by providing enhanced seeing to all science instruments.

ASMs are able to provide up to ~ 4000 actuators on secondaries ~ 1m in diameter. These would provide high-order VisAO correction on an 8-m class telescope. Current ASMs have ~ 1000 actuators for NIR correction. The Italian company Adoptica is currently the only ASM manufacturer and produced the LBT and ESO systems and are developing the ESO ELT M4, the Subaru ASM as well as the ASMs for each of the GMT sub-apertures. The Netherlands Organisation for Applied Scientific Research (TNO) is currently developing an alternative ASM design and is working with U. Hawai'i and LAO/UCO for on-sky testing at the IRTF and the UH88. Steward Observatory is also working on a modified actuator design for an Adoptica system currently undergoing testing at the MMT.

ASMs, including the facesheets, are expensive items. An Adoptica ASM (~ 1m in diameter) for an 8m telescope is ~ $5M - $8M. The facesheets are currently being produced by the French company SAGEM.

There is no US vendor for ASMs. The US Community should evaluate developing its own ASM program to remove the dependency on European sources and potentially foster technical improvements. Utilizing common technology across telescopes could be beneficial to the US observatories in the long-run and would create an industrial base for large optical elements (facesheets), actuators, and sensor technology. Such a program could involve observatory consortia, national laboratories, university departments and technical (opto-mechanical) businesses and might tie into the NSF's TIP program.

#### B.4.2. Deformable Mirrors

For the ELTs, especially for high-contrast imaging, the most pressing need is the development of large actuator count deformable mirrors (DMs) required to achieve > 80% Strehl ratios at short wavelengths. DMs with ~250 x 250 actuators, corresponding to a 12 cm spacing on a 30 m aperture. Such a DM will require a peak-to-valley stroke of up to ~ 5μm surface running at a minimum of 2kHz with a final goal of 5kHz –10kHz.

B.5. ExAO Development for ELTs.

The following apply primarily to future development of ExAO systems for ELTs but also have more general applications.

The development of future ExAO high-contrast systems for ELTs introduces a number of technological challenges in terms of instrumental hardware and algorithms and software for both control and analysis. The combination of high-order wavefront sensing and control, coronagraphy, post-coronagraph wavefront control, and post-processing must be able to characterize planets with planet-to-star contrast ratios of ~ $10^{-7}$ – $10^{-9}$ to achieve the goals of the Decadal Survey. We note that these developments will have strong connections to Habitable Worlds Observatory science and technology.

1. Detectors for Wavefront Sensing
   There will be a requirement for large format, high-speed, low-noise detectors with readout times shorter than the frame time corresponding to at least 512 x 512 pixels sampling at a minimum rate of 2 kHz, with a goal of 5 kHz to 10 kHz, and read noise less than $1e^{-}$. A global shutter will also be required with a low latency and a 100% duty cycle.

2. Wavefront Reconstruction
   Wavefront reconstruction algorithms, including especially nonlinear algorithms, need to be further developed to improve the dynamic range of the most sensitive (photon efficient) wavefront sensors. Advances in machine-learning algorithms show great promise in this area.

3. Wavefront Control
   For ExAO, the ability to suppress quasi-static speckles with post-coronagraphic wavefront sensing and correction must be developed for routine on-sky use by current instruments. Daytime and on-sky calibration techniques should be developed and optimized to correct for non-common path aberrations.

Optimized control laws, including prediction and perhaps nonlinear algorithms, need to be further developed and demonstrated in routine use on-sky. The use of sensor fusion (e.g., accelerometers, on-instrument wavefront sensing) as part of real-time wavefront control should be investigated.

4. Coronagraphs
   Coronagraph technology needs to be improved to routinely obtain inner working angles (IWAs) of < 3 λ/D with a target of 1 λ/D, and must be demonstrated on current instruments. Continued development of coronagraph optical design and manufacturing is needed to achieve this. Photonics-based coronagraphs offer the possibility of sub-λ/D IWA, and should be demonstrated on-sky.

5. Post-Processing
   The use of AO telemetry, similar to conventional PSF reconstruction, offers great promise towards minimizing and eventually eliminating speckle noise. Such algorithms need to be developed and tested using on-sky data from current instruments.

Development of improved coronagraphy, wavefront sensing, reconstruction, and control, necessary for ELT high-contrast science, should be pursued using the currently operational ExAO instruments as pathfinders and testbeds including training of the next generation of ExAO and high contrast instrument builders.

It is also worth noting that exoplanet and high-contrast detection is not just limited to ExAO systems. AO-corrected images with the Keck II ORKID camera have demonstrated 15 mas resolution at 650 nm, using shift-and-add of short exposures with no frame selection. Similar performances have been demonstrated using SHARK-VIS at the LBTO. Speckle imaging systems on both 8-m class Gemini telescopes `Alopeke at GN & Zorro at GS provide high-resolution, diffraction-limited optical imaging at visible wavelengths (FWHM~0.02" at 650 nm) which can also be used for HWO precursor observations for companion searches. Speckle imaging has a more limited dynamic range than Coronagraphic AO imaging. The use of speckle processing after partial AO correction, such as with a GLAO system, can be investigated as this will reduce the total number of atmospheric speckles and increase the SNR. And as with coronagraphic AO imaging, the development of more rigorous and sophisticated image processing techniques, likely using deconvolution from wavefront sensing approaches, is also worth investigating.

## B.6. Observing Efficiency

Observing time on 8-m & 30-m class telescopes is costly and AO operations can be problematic because of the complexity of the AO systems. Maximizing observing efficiency and up-time should be a goal of current and future AO facilities.

1. AO system ease of use, are they “turn-key” systems?

   AO system operation should be as automated as possible without the need for additional operations personnel (e.g. operable by the already present telescope operator). Normal operations like acquisition and dithering should be automatic and efficient, including LGS reacquisition and loop closure following observing closures (see Appendix §B3). Lessons can be learned from the more efficient systems.

2. Reliability.

   Hardware and software issues with the AO system can interrupt observing. These can range from causing the AO loop to open or lead to AO operations being ceased because of serious instrumental hardware problems. Hardware reliability issues also include the amount of potential AO observing time lost due to routine system maintenance, interventions to repair system faults, and the availability of spare components. Realistic engineering time estimates should be made available for facility support.

3. Robustness of the AO systems.

   AO systems should be robust against variable seeing or transparency conditions. This is expected to be more problematic with an increased number of sub-apertures. AO control systems should be designed for system robustness, an example being dynamic re-binning of sub-apertures.

4. Telescope Improvements.

   The performance of AO systems can be improved by mitigation of the following telescope / observatory induced issues:

   - Telescope Vibration Environment

     Many AO systems show residual effects from the telescope vibration environment. For example, cold-head vibrations, especially on a Cassegrain mounted instrument, can excite resonance frequencies in the telescope structure leading to residual tip-tilt signal in closed loop operation. These can be somewhat mitigated by AO control systems although they would be better mitigated by confronting the sources.

- Telescope Optics

  AO performance can be affected by the optical design and optical components of the telescope. These become more significant with the push to correction at shorter wavelengths (VisAO) and include:

  - Segmented apertures

    Phasing errors between the segments of an aperture can impact AO performance. This effect is especially detrimental to high-order wavefront correction such as VisAO ExAO.

  - Low Wind Effect (LWE)

    The low wind effect (LWE) impacts large apertures when there are pupil discontinuities, such as the secondary spiders, which are significantly cooler than the ambient air. This creates phase discontinuities of the wavefront leading to residual piston and tip-tilt aberrations, known as petal modes, and particularly impacts ExAO correction. These have been observed on 8-m class systems and could be an important issue on the US-ELTs such as the segmented GMT pupil and secondary supports of the TMT. Mitigation strategies for these need to be developed and demonstrated.

  - Optical print-through

    Some optical elements, such as M2 on Gemini N., do not produce a fully flat wavefront because of the secondary mounting.

  - AO Instrument Throughput

    One possible innovation would be to include the standard AO optical relay needed to correct wavefronts in the pupil plane; but to integrate the instrument (i.e. IFU spectrograph) into the AO system and remove the additional optical relay from the instrument.

- Local Seeing

  Turbulence within the dome affects AO performance. The AO community should collaborate to develop strategies for measuring and mitigating dome- and mirror-seeing.

A Community systematic approach to these issues should be investigated, especially for telescopes with VisAO ExAO systems.

Automated AO systems such as Robo-AO are highly reliable and robust and the community should consider what lessons can be learned from their operations.

### B.7. US AO Community and Workforce

The development of a strong and diverse US AO workforce was identified as being critical for reaching the Astro2020 science goals. This workforce involves academic positions for PhD AO scientists/astronomers, postdoctoral researchers and graduate students, as well as observatory support AO scientists and the cadre of AO engineers and technicians necessary for developing and maintaining the instrumentation. While observatory support staff tend to be on regular contracts, the university/academic staff, apart from the PIs, tend to be on temporary funding ("soft" money) who can move on to more stable positions, such as industry, leading to a loss of expertise and continuity for AO development. AO support staff at major US observatories include a high percentage of non-US trained personnel, e.g. Gemini Observatory. The shortage of US-trained AO engineers and technicians is symptomatic of a larger astronomical instrumentation staffing problem within academia where personnel rely on temporary funding.

Workforce development for AO, and astronomical instrumentation in general, could be improved by emphasizing the importance of instrumentation at the graduate school-level such as (1) instrument specific PhD programs, (2) telescope operation and support MS programs, (3) a student exchange program between university departments and observatories, and (4) national and international programs to encourage students from minority-serving and less well-funded institutes to be involved in major instrument projects.

Workforce retention can be addressed by (1) longer postdoctoral researcher positions, e.g. 5 years instead of 3 years, (2) academic institutes creating and supporting longer-term and permanent technical support positions, and (3) alternative career paths to academia, such as at national research laboratories and related industries, or other applications of AO.

It is important that the astronomical instrumentation field in general embraces Equity, Diversity and Inclusivity (EDI) and how to improve it. This involves understanding the EDI demographics related to (1) professional positions, (2) professional institution type and location, (3) professional role such as academic, support, etc., and (4) professional status such as graduate student, postdoctoral researcher, academic instructor / professor, observatory scientists, engineer, and technical support staff. Improving the EDI demographics can be tied to existing national programs and EDI programs at various related institutes.

There is no US national center for US astronomical AO. By comparison the European community has centers at ONERA and LAM in France, the Max Planck Institutes in Germany, the Italian ADONI national AO program, and Durham University's Center for Advanced Instrumentation in the UK. These are relatively well-funded centers with continuous staffing. The nearest the US have had to these was the NSF's Science and Technology Center for Adaptive Optics (CfAO) which was funded between 2000 and 2010 located at the University of California, Santa Cruz. Currently there is AO and related instrument development at a number of university departments including, but not limited to, the University of Arizona, California Institute of Technology, the University of California (Berkeley, Los Angeles, San Diego, Santa Cruz), and the University of Hawai'i. In addition, there is development work ongoing at Keck Observatory and for both US-ELT programs.

A US National AO Program to address both the workforce and coordination issues would positively influence the future for AO in the US. Such a Program could be used to encourage coordination, cooperation and collaboration between the major US AO research groups, as well as with the international community, to best develop the systems to reach the Astro2020 science goals. This would include (1) a robust university outreach program, (2) theoretical development, (3) numerical simulations, (4) laboratory and telescopes experiments, and shared testbed facilities, (5) advanced systems design and (6) a coordinated workforce development program. It would also ideally lead to using common control systems, software and hardware components between future AO systems. A National Program could also sponsor routine topical AO related workshops, similar to those sponsored by the European Community OPTICON program, as well AO Summer Schools. The CfAO legacy programs (the annual CfAO Summer School and Fall Science Retreat) continue to benefit the US AO community.

### B.8. Details of Preliminary AO Roadmap

AO can and should play multiple critical roles in achieving the science goals and top space and ground-based priorities of the Astro2020 report. Table B.1 presents a proposed timeline for this report's recommended AO milestones, by calendar year, in support of the Astro2020 goals. For reference, the current space and ground-based milestones that drive the AO milestones are provided in columns 2 and 3. Two columns are then provided for each Astro202 science theme, one for direct science support with AO and one for the development of the AO systems needed to achieve the science goals. The last two columns list the AO technology and workforce development needed to achieve Astro2020's science goals. Consideration was given to milestones that should be demonstrated to inform the 2030 and 2040 decadal reviews.

**Table B.1:** Preliminary AO Roadmap - 2023 – 2030

| CY | Relevant Planned Milestones | | Worlds and Suns in Context | | New Messengers and New Physics | | Cosmic Evolution | | AO Development | |
|---|---|---|---|---|---|---|---|---|---|---|
| | **Space and Ground** | **Ground AO Systems** | **AO Science** | **AO Systems** | **AO Science** | **AO Systems** | **AO Science** | **AO Systems** | **Technologies** | **Workforce** |
| **23** | US AO Roadmap EUCLID Launch | UH2.2m Robo-AO | | | 8-10m NIR AO TDA Science for Deep Follow-Up | | | | | |
| **24** | | WMKO KAPA/OSIRIS Operations Start | HWO Target Reconnaissance Plan | | | | | 2-4m Enhanced Seeing Demonstrator | | |
| **25** | | | | | | | | | 60 x 60 DM on 8-10m | Postdoc to Position Pipeline Established |
| **26** | Rubin Operations Start | | | Predictive Control Operational 6-10m | | | | | Super-resolution on 8-10m | |
| **27** | NGRT Launch | MAVIS on VLT | | Phasing Requirement met on 6-10m | 2-4m AO TDA Science for Rubin/Roman Follow-Up | 2-4m RoboAO Network Operational | | 2-4m Sky Coverage Demonstrator | Sensor Fusion on 8-10m | |
| **28** | E-ELT First Light | GNAO/GIRMOS Operations Start | | Quasi-static Speckle Suppression to contrast of $10^{-8}$ on-sky | | | | | Commercial 60W Laser Available | |
| **29** | HWO Architecture Decisions | | Tier A Targets Surveyed to contrast of $10^{-5}$ at 100mas | | | | 8-10m Science with ORCAS 2029-32+ | 8-10m Enhanced seeing facility | PSF-R on 8-10m | |
| **30** | **Astro2030** | | | Bandwidth Requirement met on 6-10m | | | 2-4m 5y Enhanced Seeing Surveys | | Wide Band, Fast, Low Noise Detectors Available | 2x Increase in Engineer Positions ELT Fellowships Established |

**Table B.1** (cont.): Preliminary AO Roadmap 2031 – 2040

| | | | | | | | | | | |
|---|---|---|---|---|---|---|---|---|---|---|
| **31** | | | | TMT PSI Started | | | | | | 2x Increase in Researcher Positions |
| **32** | | | | GMagAOX on GMT MODHIS on TMT | | | | NFIRAOS/IRIS on TMT | Real-time Computer Power 5X Today | |
| **33** | | | | | | | | | | |
| **34** | | | | | | | | | 3.2m TMT ASM Available | 2x Increase in Post-doc Positions |
| **35** | Probe Missions Launch | US-ELTs+AO Construction Complete | | | | | | VisAO on 8-10m with 100% Sky | 2k x 2k x 2kHz Low Noise Detectors | 2x Increase in PI Level Leaders |
| **36** | | | | | 8-10m VisAO TDA Science | | 8-10m VisAO Science | | | |
| **37** | | | | | | | | | | |
| **38** | | | | | | | | | | |
| **39** | | | | PSI on TMT | | | | | | |
| **40** | **Astro2040** | US-ELTs AO 2nd Generation Performance | | | | | | | | |

B.9. Laser Communications

It should also be noted that advances in AO have enabled NASA to explore laser communications (LaserCom) as an alternative to radio communication for two-way data transmission from ground-to-space.  IR laser communications allow for space missions to increase the data in a single transmission by factors of 10x – 100x. The Laser Communications Relay Demonstration (LCRD) experiment has already demonstrated significant advantages over traditional radio communication systems. With the advent of new space missions, both manned and unmanned, it is important that this development continues and it is expected that technical advances for ground-based AO systems will also be applicable for LaserCom AO.

## C. Astronomical Adaptive Optics

### C.1. Background and History

Adaptive Optics (AO) mitigates image degradation due to atmospheric turbulence such that ground-based telescopes are able to reach their theoretical diffraction-limit as opposed to being limited by the site seeing. A typical seeing-limited image has a full-width at half maximum (FWHM) of ~ 1 arcsecond. This is site dependent and at excellent seeing sites such as Maunakea the seeing can reach ~ 0.5 arcseconds or better. However this is still substantially larger than the diffraction-limits of large telescopes. For example the diffraction-limit in the R-band (658 nm) on a 4m telescope is 0.034 arcseconds, on a 10m telescope is 0.014 arcseconds, and on a 30m telescope is 0.005 arcseconds. In the K-band (2.2 μm) for the same three apertures the diffraction-limits are 0.113, 0.045, and 0.015 arcseconds respectively. Large telescope diffraction-limits can be reached with AO systems enabling unprecedented science for ground-based observatories which complement space-based observatory data.

AO was first developed and tested for astronomical applications in the 1980s and the first AO systems came into regular operation during the 1990s. The early systems used natural guide stars (NGS), i.e. an unresolved celestial source for the wavefront sensing, but the required brightness limited the sky coverage. Laser guide stars (LGS) were developed for astronomical use in the 1990s. They became standard on large telescope AO systems in the 2000s. The LGS is an artificial beacon which is used in *lieu* of the NGS and yields significantly expanded sky coverage as they can be used at any target. However, the sky coverage is still limited because an NGS is still required for atmospheric jitter measurements. AO systems are now *de rigeur* on many astronomical telescopes of varying aperture sizes which feed a variety of science instrumentation including imagers, spectrographs and IFUs.

Astronomical AO has demonstrated its effectiveness at every major optical/infrared (OIR) observatory over the last 25 years. For the US community, this includes the current large telescopes (hereafter referred to as the 8-m class) such as both Keck I and Keck II, both Gemini telescopes, the Large Binocular Telescope (LBT), as well as the Magellan Clay telescope and the MMT. AO is also in regular use on smaller aperture telescopes (hereafter referred to as the 4-m class) such as the Palomar Hale 5m, the Lick Shane 3m and the 4.1m SOAR telescope in Chile. Additionally the Robo-AO automated survey mode has been used on smaller apertures, the 2-m class telescopes, with a new system to be recently implemented on the UH88 on Maunakea.

AO technology is continually improving with the introduction of low-noise wavefront sensor (WFS) detectors, faster computer systems for controllers, and higher-density deformable mirrors (DMs) which reduce the wavefront error on the corrected beam. Adaptive Secondary Mirrors (ASM), i.e. a DM replacing the standard telescope secondary mirror, are in operation at the Large Binocular Telescope (LBT) and UT4, one of the four ESO VLT 8-m class telescopes and one is under development for the NAOJ Subaru 8.2m.

During the past quarter century, AO has benefited and enabled many areas of astrophysics, including Solar Physics, Solar System planetary science, dynamical studies of low and mid-redshift galaxies, exoplanets, stellar populations, as well as the discovery and confirmation of the black hole in our own galaxy. As such it has produced a large number of high-impact papers and science. AO is an essential component for the next generation of Extremely Large Telescopes (ELTs) which have been designed around AO systems making use of both the large collecting area and the corresponding high angular resolution.

In short, AO is a well-established observational technique for ground-based astronomy and astrophysics, but has not yet reached its full potential at all wavelengths and over the entire sky.

### C.2. 2008 US AO Roadmap

US AO developments over the last 15 years have been influenced and motivated by the last AO roadmap,*“A Roadmap for the Development of United States Astronomical Adaptive Optics”* published in April 2008, which was sponsored by the NSF’s Division of Astronomical Sciences in response to the AO Community’s request for a reliable public funding source for AO system research, development, and implementation.

Given that the last AO roadmap was formulated a decade and a half ago, the US AO community have been looking towards what future AO systems will be needed to either replace and/or augment the existing ones in light of the science goals laid out by Astro2020. This takes account of the changing astronomical landscape over the next decade including the commissioning and routine operation of the Vera C. Rubin Observatory, the 30-m class “telescopes being completed and commissioned, and a number of planned and proposed space missions such as the Habitable Worlds Observatory (HWO) and the Nancy Grace Roman Observatory.

### C.3. Adaptive Optics Performance

The actual performance of an AO system is defined by how well the atmospherically-induced wavefront aberrations have been corrected. This depends upon a number of factors relating to the components of the AO system.

The wavefront aberration is measured by the wavefront sensor (WFS). How well it is measured depends upon the number of sub-apertures and their size relative to the seeing typically defined by the Fried parameter $r_0$, the number of WFSs being used, the type of guide star being used (NGS or LGS), and the measured SNR on the WFS which depends upon the sensor's sensitivity and noise statistics, the brightness of the guide star, and the integration time.

The wavefront correction is determined by the deformable mirror(s) being used. More specifically the number of actuators, and for some systems the number of mirrors and their conjugation altitudes. The final wavefront correction also depends upon the speed at which the correction takes place which is determined by the correlation time of the atmosphere, $\tau_0$.

From an operational perspective, the AO Performance is typically determined using several figures of merit. These are: (1) the Image Quality which can be measured by a number of metrics; Strehl Ratio, Encircled Energy in a fixed aperture, and Contrast; (2) the Sky Coverage which is the fraction of sky which can be observed at fixed image quality (Strehl ratio). The sky coverage depends upon the availability of suitable NGSs for either full correction or for tip-tilt correction when using LGSs; (3) the bandwidth over which the corrected wavefront is available for science; and (4) the field-of-view (FoV) of the corrected field which determines the type of science which are available.

## C.4. Adaptive Optics Systems

A suite of different AO systems which have been developed for different applications, ranging from very narrow-field, a few arcseconds, high Strehl ratio extreme AO systems to wide-field, several arcminutes, partial AO correction.  The vast majority produce corrected beams primarily at NIR wavelengths although some are producing correction in the visible. The following is a list of the type of AO systems currently available to the US community.

### C.4.1. Single-Conjugate Adaptive Optics (SCAO)

In a single-conjugate adaptive optics (SCAO) system, a single  guide star is used to measure the wavefront phase. This can be done with either a natural guide star (NGS) or a laser guide star (LGS). Generally the deformable mirror (DM) is conjugated to the ground. These were the first AO systems to be installed on astronomical telescopes and have become the workhorse facilities feeding a variety of science instruments on various telescopes.  SCAO systems typically provide diffraction-limited (DL) performance over a small field-of-view (FoV) ~ 5 - 10 arcseconds and the vast majority of them provide correction at NIR wavelengths.

The performance of these systems is limited by the brightness of the guide star. They perform well on bright NGSs and the performance degrades as NGS brightness decreases due to reduced flux on the wavefront sensor (WFS). As such they have limited sky coverage. Increased sky coverage is obtained by using an artificial reference source, an LGS beacon, which can be pointed anywhere in the sky. The sky coverage then becomes limited by the availability of an NGS used to measure the tip-tilt mode not available from the LGS. However, there is a performance loss due to focal anisoplanatism. This is due to the fact that the LGS is situated at a finite distance from the telescope aperture. An NGS measures the wavefront distortion in a cylinder above the telescope pupil whereas an LGS measures the distortion within a cone with the LGS at its apex thus leaving unsensed wavefront disturbances from the astronomical target.. This focal anisoplanatism is worse for Rayleigh beacons which have a typical range of 15 km compared to the Sodium guide stars which create a beacon at an altitude of ~ 100km.

### C.4.2. Multi-Conjugate Adaptive Optics (MCAO)

As noted above SCAO has a narrow FoV. One approach to overcome this is by using a multi conjugate adaptive optics (MCAO) system. This makes use of multiple DMs to measure and correct the wavefront aberrations for different layers in the atmosphere such that each DM is conjugated to a different atmospheric layer. Multiple reference sources are necessary to measure the wavefront in different directions from the on-axis pointing and a tomographic reconstruction of the atmosphere yields DL performance over a wider FoV ~ 1 - 2 arcminutes.

### C.4.3. High-Contrast Extreme Adaptive Optics (ExAO)

Extreme AO systems are designed for high-contrast imaging and are primarily, but not exclusively, used for exoplanet imaging typically using coronagraphic systems to reduce the flux from the parent star. Such direct imaging explores a different “phase space” of exoplanet detection than either radial velocity or transit techniques. These are NGS systems using the bright target star for the reference source.

The first ExAO systems provided high-contrast observations at near-infrared wavelengths looking for younger planets via their thermal signatures. More recent systems have been designed for improved contrast permitting visible light imaging in reflected starlight thereby enabling detection of different classes of exoplanets.

### C.4.4. Enhanced-Seeing / Ground-Layer Adaptive Optics (GLAO)

GLAO systems provide improved image quality over natural seeing but do not reach the DL. They typically provide enhanced or “super seeing" which decreases the natural seeing FWHM of 0.6 – 1 arcseconds to ~ 0.2 – 0.4 arcseconds depending very much on the atmospheric turbulence profile of the site. There are also other approaches, which decrease the seeing FWHM, that fall into this category. Both are discussed below.

GLAO systems provide partial correction of atmospheric blurring over a significantly large field of view (~ 10 arcminutes) and over a broader wavelength range (optical and infrared) than classical AO systems. Existing seeing-limited instruments can be used to take advantage of this improved image quality and correspondingly reduced source confusion to improve science productivity. The applicable science is very broad and includes extra-galactic spectroscopic surveys, intergalactic and circumgalactic medium studies, stellar systems, stellar population studies in the Milky Way and nearby galaxies, as well as solar system studies.

GLAO systems are more effective when the atmospheric turbulence is typically dominated by a significant, but thin, ground layer such as that which has been measured at Maunakea although they can be used, but not as effectively, under differing atmospheric profiles. The dominant ground-layer allows for uniform correction over a wide field of several arcminutes. To achieve this, a GLAO system uses multiple guide stars, typically LGS, distributed over a field of a few arcminutes, and uses the average measured wavefront for correcting the DM.

Non-GLAO enhanced seeing AO systems are also in operation. These provide low-order correction over smaller field sizes than GLAO, and do not differentiate the atmospheric layers.

#### C.4.5. Laser Tomographic Adaptive Optics (LTAO)

The focal anisoplanatism limitations of LGS SCAO systems can be mitigated by using multiple LGSs to completely sample the atmosphere in the cylinder above the telescope. This increases the higher performance sky coverage, and like the GLAO systems discussed above are limited by the available TT NGSS. The wavefront phases in the direction of the object of the target are obtained through a tomographic reconstruction process and lead to high AO performance. However, the FoV is limited, as for SCAO, because a single wavefront corrector is used.

#### C.4.6. Multi-Object Adaptive Optics (MOAO)

MOAO is an alternative to the wide field correction of MCAO. It corrects in the direction of the targets of interest as opposed to correcting the whole field. The only MOAO system which will be available to the US community is the near future will be the Gemini Infra-Red Multi-Object Spectrograph (GIRMOS). This is a four-arm MOAO IFU spectrograph which will interface with the GNAO system at Gemini N. It will be used with the GNAO wide-field mode which will provide improved image quality, and will be able to select targets within the 2 arcminute FoV. The MOAO will use small DMs operating in open-loop to achieve near-DL performance for a 3” x 3” FoV for each of the four target. Commissioning is expected in 2027.

### C.4.7. Autonomous Survey Adaptive Optics (Robo-AO)

Robo-AO is an autonomous combined LGS adaptive optics system and science instrument. The system delivers near DL performance at visible and NIR wavelengths and is capable of executing large-scale surveys, monitoring long-term astrophysical dynamics, and characterizing newly discovered transients. Robo-AO uses a UV Rayleigh LGS (355 nm) in order to mitigate laser conflicts with aircraft they are exempt from FAA requirements, and have worked with the US Space Command’s Laser ClearingHouse to develop a unique approach for dealing with potential satellite conflicts. An upgraded version, Robo-AO-2, is currently installed and ready for operation at the UH88 on Maunakea.

Table C.1 Breaks down the general specifications for the different AO Systems.

**Table C.1**: Overview of the different types of AO Systems.

| AO System | Science Wavelength Range | Field of view (diameter) | Image Quality / Figure of Merit | Sky Coverage |
|---|---|---|---|---|
| **SCAO (NGS)** | 0.6 $\mu$m - 5 $\mu$m | 10" - 20" | DL in NIR<br>DL FWHM in visible | Few percent |
| **SCAO (LGS)** | I - K Band | 20" | Modest Strehl in NIR | 50% |
| **LTAO (LGS)** | I - K Band | 21" | DL in NIR<br>DL FWHM in visible | 50% |
| **ExAO / High Contrast** | 0.5 $\mu$m to 2.2 $\mu$m | 1" - 5" | DL Red -- NIR | Individual targets |
| **Ground-Layer (GLAO)** | Entire visible/IR | ~ 5' - 15' | Enhanced Seeing (FWHM ~0.4" in visible) | 100% |
| **Enhanced Seeing (ES)** | Entire visible/IR | 1' - 2' | Enhanced Seeing (FWHM ~0.4" in visible) | 100% |
| **Multi-Conjugate (MCAO)** | Visible/NIR | 1' - 2' | Modest Strehl in NIR | 50%+ |
| **Multi-Object (MOAO)** | NIR | 2' - 5' | Modest Strehl in NIR | 50%+ |
| **Autonomous Survey Adaptive Optics (Robo-AO) (LGS)** | Visible/NIR | 30" | Low/Modest Strehl | 50%+ |

## D. Adaptive Optics Telescopes available to the US Community

It's important to remember that the US OIR system is a Public-Private partnership. The Public side is represented by the Observatories which receive direct operational funding from the Federal coffers, more specifically via the National Science Foundation's (NSF) Division of Astronomical Sciences (AST). These observatories also have domestic and international partners. The Private Observatories receive their operational funding via various Research Institutes, Universities, Foundations, and non-US partners. They also receive some funding from Federal coffers for specific projects, typically via competitive NSF or NASA programs.

Two of these observatories are general community access operated by AURA under a cooperative agreement with the NSF and under the auspices of NSF's NOIRLab. The first of these is the Gemini Observatory which operates two 8.1m telescopes, one in the Northern Hemisphere on Maunakea in Hawai'i, and the other in the Southern Hemisphere on Cerro Pachón in Chile. The NSF's NOIRLab partnership entails access to the full US astronomical community dependent upon peer-reviewed observing proposals. Gemini is an international partnership with the United States as the primary partner with the other partners being Canada, Brazil, Argentina, and Korea. The two hosts, U. Hawai'i (GN) and Chile (GS) also have an allocation of observing time. The second is the the 4.1-meter Southern Astrophysical Research (SOAR) Telescope located nearby the Gemini S. Telescope on Cerro Pachón and is a joint project of the Ministério da Ciência, Tecnologia e Inovações do Brasil (MCTIC/LNA), NSF's NOIRLab, the University of North Carolina at Chapel Hill (UNC), and Michigan State University (MSU).

The private access observatories offer AO access to various members of the US astronomical community with connections to their own operational and funding sources.

WMKO is managed by California Association for Research in Astronomy (CARA) representing the California Institute of Technology (CIT) and the University of California (UC). As with Gemini, the U. Hawai'i is the host, and WMKO receives NASA funding giving US Community access.

LBTO is operated by an international consortium comprising two international partners and two domestic partners. The international partners are (1) the Italian National Institute for Astrophysics (INAF) which consists of twenty separate research facilities, (2) a German consortium of five research institutes, the LBT Beteiligungsgesellschaft, composed of (a) the Max-Planck-Institute for Astronomy in Heidelberg, (b) the Max-Planck-Institute for Extraterrestrial Physics in Garching, (c) the Max-Planck-Institute for Radio Astronomy in Bonn, (d) the Leibniz-Institute for astrophysics in Potsdam, and (e) the Landessternwarte Heidelberg (LSW). The two US partners, which account for 50% (25% each) of the allocated observing time are (1) the University of Arizona (UA), which coordinates observing for the three Arizona State Universities, (a) UA, (b) Northern Arizona University (NAU), and (c) Arizona State University (ASU), and (2) The Ohio State University (OSU) which coordinates the participation of (a) OSU, (b) the University of Notre Dame (UND), (c) the University of Minnesota (UMN), and (d) the University of Virginia (UVA).

The Magellan Telescopes are a pair of 6.5 m diameter telescopes located at the Las Campanas Observatory in Chile. They are operated by a consortium consisting of the Carnegie Institution for Science, the University of Arizona, Harvard University, the University of Michigan and the Massachusetts Institute of Technology. The Clay Telescope (aka Magellan II) has the extreme AO system, MagAO-X which is available to any Magellan partner institute.

The MMT, located on the site of the Fred Lawrence Whipple Observatory located on Mount Hopkins, Arizona, also has a 6.5 m aperture. It is operated by the MMT Observatory which is a joint venture of the Smithsonian Institution and the University of Arizona. Currently the MMT Adaptive Secondary Mirror is being upgraded as part of the MAPS project.

The 5m Hale Telescope, located at Palomar Observatory in Southern California, is operated by Caltech Optical Observatories for the benefit of Caltech, Jet Propulsion Laboratory, Yale University, and NIAOC. The Hale is outfitted with the PALMAO adaptive optics facility instrument, a natural guide star AO instrument supporting multiple exchangeable back-end instruments. Beginning in 2025, the Hale will also host SIGHT LGS enhanced seeing facility, currently in development with NSF and Mt. Cuba Astronomical Foundation support. SIGHT will be integrated into the telescope above the Cassegrain focus and improve seeing for the Hale's workhorse single-object spectrographs.

The 3m Shane Telescope is located at Lick Observatory on Mt. Hamilton in California and is part of the University of California Observatories (UCO) serving the UC community. The UH88 is a 2.2m telescope situated on Maunakea and serves the University of Hawai'i (UH) community. In addition to the US observatories listed above, US astronomers also have access to the NAOJ Subaru 8.2m on Maunakea thanks to the Exchange programs Subaru has with both Gemini Observatory and WMKO.

Table D.1 provides a summary of the AO enabled Observatories and their partner institutes. Note that the percentage of time available to the partners is not included.

| Institutes | Gemini | WMKO | LBTO | Mag. | MMT | Hale | SOAR | Lick Shane | UH88 | Subaru |
|---|---|---|---|---|---|---|---|---|---|---|
| US Community | NSF | NASA | | | | | NSF | | | Gemini |
| Arizona State U. | | | x | | | | | | | |
| California Institute of Technology | | x | | | | x | | | | WMKO |
| Carnegie | | | | x | | | | | | |
| Harvard Smithsonian | | | | x | x | | | | | |
| Jet Propulsion Laboratory | | | | | | x | | | | |
| Mass. Institute of Technology | | | | x | | | | | | |
| Michigan State U. | | | | | | | x | | | |
| Northern Arizona U. | | | x | | | | | | | |
| Northwestern U. | | x | | | | | | | | |
| Ohio State U. | | | x | | | | | | | |
| U. Arizona | | | x | x | x | | | | | |
| U. California | | x | | | | | | x | | WMKO |
| U. Hawai'i | x | x | | | | | | | x | x |
| U. Michigan | | | | x | | | | | | |
| U. Minnesota | | | x | | | | | | | |
| U. North Carolina | | | | | | | x | | | |
| U. Notre Dame | | x | x | | | | | | | |
| U. Virginia | | | x | | | | | | | |
| Yale U. | | x | | | | x | | | | |

**Table D.1:** The AO enabled telescopes and the US institutional community access to each of them. General US Community access to Gemini & SOAR is via the NSF partnership and general US Community access to WMKO is via NASA. Both the WMKO and Gemini communities have access to Subaru via Exchange Programs with each Observatory. Note that SOAR is available to the whole US Community but only the US partner institutes are included.

Missing from this list are (1) the AO facilities used by the Solar Physics community: the VisAO MCAO systems routinely in use at the 1.6m clear aperture at BBSO in California and at the 4m DKIST on Maui in Hawai'i', and (2) the AO facilities for the two US ELP projects, TMT and GMT. Both US-ELT projects are international. For the TMT the US partners are the California Institute of Technology and the University of California. The international partners are (1) the Department of Science and Technology (DST) of India, (2) the National Institutes of Natural Sciences incorporating the National Astronomical Observatory of Japan (NINS), and (3) the National Research Council (NRC) of Canada. For the GMT the US Partners are (1) the Carnegie Observatories, (2) the Harvard-Smithsonian Center for Astrophysics, Texas A&M and the University of Texas at Austin, (3) the University of Arizona, and (4) the University of Chicago. The international partners are (1) Astronomy Australia Limited and the Australian National University, and (2) the Korean Astronomy and Space Science Institute.

# E. The Current Adaptive Optics Landscape

## E.1. Current Adaptive Optics Systems for the US Community

The US Community currently has access to the following AO-equipped Observatories, including one non-US observatory via exchange programs. Note that the MMT is not currently included as their ASM is still undergoing testing:

- Gemini Observatory (8m)
- W.M. Keck Observatory (10m)
- Large Binocular Telescope Observatory (2 x 8.4m)
- Magellan Telescope (6.5m)
- Palomar Observatory (5m)
- SOAR (4m)
- Lick Observatory - Shane (3m)
- U. Hawai'i UH88 (2.2m)
- NAOJ-Subaru (via exchange time with UH, Gemini and WMKO) (8m)

The following are descriptions of the currently and soon-to-be AO systems available to the US community as outlined in Table 3 in the main text.

SCAO systems are currently in operation at Gemini N., both Kecks, LBT, Hale, & Shane. LBT uses an adaptive secondary mirror (ASM) as the AO system's DM which has the potential of a wider FoV and greater efficiency with fewer reflecting surfaces as well as permitting observations in the thermal IR. The University of Arizona are currently testing a new ASM actuator design at the MMT with the expectation that the system will become a facility instrument within the next couple of years. The Subaru telescope currently uses a 188 actuator DM but is upgrading to a thousand plus actuator ASM which is expected to see first light in 2024. UH is currently investigating a different ASM design for testing at the NASA IRTF.

Currently, the only available MCAO system for the US Community is the GeMS system at Gemini S. which has been in operation for the last decade. This uses sodium LGS beacons for the wavefront sensing and the wavefront correction is currently accomplished using two DMs conjugated to altitudes of 0 km and 9 km. An NGS MCAO system has been demonstrated at the LBT (LINC-NIRVANA) which was developed by its German and Italian partners but at the moment it is unlikely it will become a facility instrument and will not be available.

High Contrast systems have been available for the past decade with GPI on Gemini S. and MagAO at Magellan. Both have now been replaced with a GPI upgrade (GPI 2.0) scheduled for Gemini N. in 2024 and MagAO-X is already in operation at Magellan. SCxAO has been under development during this time making use of the Subaru AO188 corrected beam to feed it. The US Community has access to SCExAO via the Keck and Gemini exchange time as well as separate UH access. During 2024 the two SHARKS at LBT will be commissioned and will be available as PI instruments for the LBT Community. These provide high-contrast imaging independently at both visible and NIR wavelengths and as they use different sides of the LBT, observing could be used simultaneously. Keck currently has KPIC which is an upgrade to the Keck II telescope's adaptive optics system and instrument suite using vortex coronagraphic masks to NIRC2 and a new infrared pyramid wavefront sensor. HAKA, the Keck II high-order AO system, is expected to be available by 2027 for NIR and visible observations.

SAM (SOAR Adaptive Optics Module) uses a UV (355 nm) range-gated Rayleigh beacon to select the low altitude turbulence to obtain 0.3" or 0.4" over a 3-arcmin square field. This is currently being upgraded to SAMplus to enhance its performance at visible wavelengths for both imaging and spectroscopy use. The ARGOS GLAO system at LBT is also a Rayleigh system using 532 nm lasers and has been commissioned on both sides of the telescope. It is currently not operational and is being reevaluated ARGOS makes use of the ASMs at the LBT and the Subaru ULTIMATE GLAO system will also make use of an ASM to take advantage of their 14 arcminute FoV and is expected to be in operation by 2027.

The ‘imaka system using multiple NGSs was a test-bed system on the UH88 giving improved and stable FWHM across ~ 11 arcminute FoV. The ‘imaka team are working with a different ASM design to be tested on the NASA IRTF on Maunakea. SIGHT is being developed for use at the Hale using UV LGS and will support multiple cassegrain spectroscopic instruments. First light is expected in 2024. The GNAO system currently d=being developed for Gemini N. will also have a 2 arcminute wide-field GLAO mode.

Two non-GLAO enhanced seeing AO systems are currently in operation. The first of these is at Gemini N.. The Altair optical design limits the sky coverage because the NGS TT has an FoV ~ 25". By using the peripheral wavefront sensor (P1) the TT sky coverage increases but with decreased, i.e. non DL, resolution yielding a FWHM ~ 0.1" - 0.2" suitable for non-imaging spectrographic and integral field unit observations. The second is the pair of SOUL / LUCI systems at the LBT. By reducing the number of corrected modes to 11, partial correction with a FWHM ~ 0.2" over a 4 arcminute FoV has been obtained and is routinely used for the LUCI imager / spectrograph.

It’s also important to note that both Gemini and WMKO are currently considering GLAO systems using ASMs for the future

There are no LTAO systems currently available to the US community although there are three being developed for the 8-m class telescopes on Maunakea. GNAO is being developed for Gemini North and is expected to see first light and commissioning in 2028. It will replace the current Altair SCAO facility. KAPA is being developed for Keck1 with first light expected in 2024 and the START system is being developed for Subaru making use of their ASM. It is noted that GNAO will feed GIRMOS which will be the first MOAO system in use by the US community.

Table E.1 provides a list by AO type of the science capabilities currently available or soon to be available to the US community. Of note is the growth in the number of AO systems over the last couple of decades with the addition of high-contrast and GLAO type systems and the development of improved performance, general purpose workhorse AO such as the soon-to-be-implemented LTAO systems. Also of note is the single MCAO system currently in operation and only a single MOAO is in development, both at Gemini Observatory. As can be seen from the table, there is a wide variety of AO systems available to the US community on multiple telescopes.

**Table E.1:** AO systems available to the US community.

| AO System Category | Telescope | Aperture (m) | AO System | NGS | LGS | LGS $\lambda$ (nm) | Instrument | $\lambda$ (μm) |
|---|---|---|---|---|---|---|---|---|
| SCAO - Facility | Gemini N. | 8.1 | Altair | x | x | 589 | NIRI | 1 - 5 |
| | | | | | | | GNIRS | 0.8 - 5.4 |
| | | | | | | | NIFS | 0.95 - 2.4 |
| | Keck 1 | 10 | K1AO | x | x | 589 | OSIRIS | 0.95 - 2.4 |
| | Keck 2 | 10 | K2AO | x | x | 589 | NIRC2 | 0.95 - 5 |
| | | | | | | | NIRSPEC | 0.95 - 5.5 |
| | LBT | 2 x 8.4 | SOUL / LUCI | x | | | LUCI | 0.9 - 2.4 |
| | Subaru | 8.2 | AO188 | x | x | 589 | IRCS | 0.95−4.80 |
| | | | | | | | 3DII | 0.36 - 0.9 |
| | Palomar | 5 | PALMAO | x | | | PHARO | 1 - 2.5 |
| | | | | | | | PARVI | 1.2 - 1.8 |
| | SHANE | 3 | Shane | x | x | 589 | ShARCS | 0.9 - 2.5 |
| | MMT | 6.5 | MAPS | x | | | ARIES | 1 - 2.5 |
| | | | | | | | MIRAC 5 | 2-13 |
| MCAO - Facility | Gemini S. | 8.1 | GeMS | | x | 589 | GSAOI | 0.9 - 2.4 |
| | | | | | | | Flamingos 2 (2025) | 0.95 - 2.4 |
| | LBT | 8.4 | LINC-NIRVANA | x | | | LINC-NIRVANA | 1 - 2.5 |

| AO System Category | Telescope | Aperture (m) | AO System | NGS | LGS | LGS λ (nm) | Instrument | λ (µm) |
|---|---|---|---|---|---|---|---|---|
| **High Contrast Extreme AO** | **Magellan** | 6.5 | MagAO-X | x | | | EMCCD | 0.5 - 1 |
| | | | | | | | VIS-X | |
| | **Subaru** | 8.2 | SCExAO / AO 188 | x | | | CHARIS | 1.1 - 2.4 |
| | | | | | | | VAMPIRES | 0.6 - 0.8 |
| | | | | | | | FPDI | 0.95-1.86 |
| | | | | | | | MEC | 1.0 - 1.2 |
| | | | | | | | REACH / IRD | 0.97-1.75 |
| | **LBT** | 8.4 | SOUL / LBTI | x | | | SHARK-VIS (2024b) | 0.4 - 0.9 |
| | | | | | | | SHARK-NIR (2024a) | 0.96 - 1.7 |
| | | 2 x 8.4 | SOUL / LBTI | x | | | LMIRCam | 1 - 5 |
| | | | | | | | NOMIC | 8 - 13 |
| | **Gemini N.** | 8.1 | GPI 2.0 (2024) | x | | | IFS / Polarimeter | 0.95 - 2.4 |
| | **Keck 2** | 10 | K2AO+PyWFS | x | | | KPIC + NIRSPEC | 1.5 - 3.5 |
| | | | HAKA (1027) | x | | | SCALES (2025) | 1 - 5 |
| | | | | | | | KPIC + NIRSPEC | 1.5 - 3.5 |
| | | | | | | | HISPEC (2026) | 0.95 - 2.5 |
| **Enhanced Seeing / Ground Layer AO** | **LBT** | 8.4 | ARGOS | | x | 532 | LUCI | 0.9 - 2.4 |
| | | | SOUL / LUCI | x | | | LUCI | 0.9 - 2.4 |
| | **UH88** | 2.1 | Imaka | x | | | 11' x 11' CCD | 0.35 - 1.1 |
| | **Palomar** | 5 | SIGHT | | x | 355 | NGPS | 0.36 - 2.5 |
| | **SOAR** | 4.1 | SAMplus (2024) | | x | 355 | SIFS | 0.4 - 0.8 |
| | | | | | | | SAMOS | 0.35 - 0.9 |
| | **Subaru** | 8.2 | ULTIMATE (2027) | | x | 589 | MOIRCS (Phase I) | 0.9 - 2.5 |
| **Laser Tomographic AO** | **Gemini N.** | 8.1 | GNAO (2028) | | x | 589 | GIRMOS | 1 - 2.4 |
| | **Keck 1** | 10 | KAPA (2024) | | x | 589 | OSIRIS | 0.95 - 2.4 |
| | | | | | | | LIGER (2027) | 0.8 - 2.4 |
| | **Subaru** | 8.2 | ULTIMATE-START | | x | 589 | IRCS | 0.95–4.80 |
| | | | | | | | 3DII | 0.36 - 0.9 |
| **Autonomous Survey AO** | **UH88** | 2.1 | Robo-AO-2 | | x | 355 | Robo-AO-2 | 0.5 - 1.7 |

| Demonstrator | Development |
|---|---|

### E.2. Adaptive Optics Systems for the US-ELTs

There are also first generation AO systems under development for the two US Extremely Large Telescopes which includes both the Giant Magellan Telescope (GMT) and the Thirty Meter Telescope (TMT).

The GMT AO system is designed to be an integral part of the telescope which provides LGSs, WFSs, with AO corrected wavefronts to every currently planned GMT instrument. There are currently three first generation planned AO observing modes. These are (1) Natural Guidestar (NGAO), (2) Laser Tomography (LTAO), and (3) Ground Layer AO (GLAO). All three will use a segmented ASM for the wavefront correction. The NGAO mode will provide extreme AO performance using bright guidestars. The LTAO mode will use six lasers to improve the sky coverage but with poorer AO performance than the NGAO mode. And the GLAO mode will use four NGS and it is expected that it will be used as frequently as possible depending upon suitable observing conditions,

The Narrow Field InfraRed Adaptive Optics System (NFIRAOS) is the TMT's first light LGS MCAO system designed to provide diffraction-limited imaging over a 34 x 34 arcsecond high sky coverage in the NIR (J, H, and K) for the first-light imager IRIS and spectrograph MODHIS. It also has a smaller-field NGS capability. Both the SCAO and MCAO modes are expected to have a 70% Strehl ratio performance.

### E.3. The Non-US Global Adaptive Optics Landscape

The two major Observatories using Adaptive Optics outside of the US are the Japanese Subaru Observatory on Maunakea, HI operated by NAOJ and the VLT facility at Paranal, Chile operated by the European Southern Observatory. The Subaru systems have been already discussed as the US community have access via the Gemini and Keck exchange programs.

#### E.3.1. ESO VLT Adaptive Optics Systems

The European Southern Observatory has a suite of four 8.2m telescopes at Paranal in northern Chile. Two of the four telescopes, UT3 & UT4, support adaptive optics systems and UT4 is a dedicated Adaptive Optics Facility (AOF) with a 1170-actuator ASM and four LGSs. There are four AO systems implemented at UT4. These are (1) GALACSI, (2) GRAAL, (3) ERIS, and (4) MAVIS. The high contrast SPHERE AO system is available on UT3.

GALACSI has two different AO modes. In its wide field mode it performs as a GLAO system with a 1' x 1' FoV. In its narrow field mode it performs as an LTAO system over a 7.4" x 7.4" Fov with moderate Strehl performance. GALACSI feeds the MUSE IFUs at visible wavelengths (0.4μm – 0.9μm).

GRAAL is a GLAO system using the 4 LGSs and a single NGS to provide correction over a 7.5' x 7.5' FoV with a FWHM of ~ 0.1 – 0.2”. It feeds HAWK-I, an NIR (0.85 µm – 2.5 µm ) wide-field imager.

ERIS provides NIR diffraction limited imaging at 1 µm – 5 µm over fields of almost 30″ and 60″, integral field spectroscopy at 1 µm – 2.5 µm at resolutions of R ~ 5000 and R ~ 10 000 over a field of view from 0.8″ to 8″, high-contrast imaging with focal and pupil plane masks, and sparse aperture masks; and long-slit spectroscopy at 3 µm – 4 µm. It’s AO provides high-order correction using a natural guide star (NGS) or a laser guide star (LGS) with a tip-tilt star and it also enables an enhanced seeing mode and seeing-limited observations.

SPHERE, on UT3, is the extreme adaptive optics system and coronagraphic facility at the VLT. Its primary science goal is imaging, low-resolution spectroscopic, and polarimetric characterization of extra-solar planetary systems at optical and near-infrared wavelengths. It operates in the NIR from 0.95 µm – 2.32 µm, depending upon the science instrument.

E.3.2. ESO ELT Adaptive Optics Systems

The ESO Extremely Large Telescope (ELT) is currently under construction in Chile. Its AO systems are built around the large (2.4m) deformable M4 mirror which comprises 5000 actuators. M4 provides SCAO (NGS and LGS), LTAO, and GLAO AO modes with up to eight LGSs. An MCAO mode is available using the post M4 AO module MORFEO, previously called MAORY. The AO systems feed the following instruments: HARMONI, MICADO, METIS, ANDES, & MOSAIC.

HARMONI (the High Angular Resolution Monolithic Optical and Near-infrared Integral field spectrograph) is the visible and near-infrared integral field spectrograph and will be compatible with SCAO & LTAO modes and will also have high-contrast capability. Its spectral resolution will range from R~3000 to R~17000 with a range of spaxal scales.

MICADO (the Multi-AO Imaging Camera for Deep Observations) provides for diffraction-limited imaging and long-slit spectroscopy at near-infrared wavelengths. It has a field of view of 50 arcseconds at 0.8µm – 2.4 µm, 50 microarcseconds astrometric precision for brighter sources), focal and pupil plane coronagraphs, and slit spectroscopy from 1.49 µm – 2.45 µm and 0.82 µm – 1.55 µm at R ~ 20K.

METIS (the Mid-infrared ELT Imager and Spectrograph) will enable high contrast imaging and integral field unit (IFU) spectroscopy (R~100,000) in the thermal- and mid-infrared (3 µm – 13 µm). It will proved (1) Direct imaging in the L, M and N bands,

(2) High-contrast imaging (HCI) in the L, M and N bands, (3) Long-slit spectroscopy in the L, M and N bands, (4) IFU spectroscopy in the L and M bands, and (5) IFU spectroscopy combined with coronagraphy.

ANDES (the ArmazoNes high Dispersion Echelle Spectrograph), and formerly called HIRES, is a multi-purpose spectrograph at visible to near-infrared wavelengths. It consists of three fiber-fed spectrographs (UBV, RIZ, YJH) with R~100,000 covering 0.4 µm – 1.8 µm with a K band potential upgrade. It operates in seeing- and diffraction-limited (SCAO) modes.

MOSAIC is a Multi-Object Spectrograph covering the Visible and Near Infrared bandwidth (0.45 µm – 1.8 µm) with two observing modes: spatially resolved spectroscopy with 8 integral field units; and the simultaneous observation of 200 objects in the VIS and NIR.. In its High-definition mode (HDM) it will permit simultaneous observations of eight integral field units (IFUs) deployed within a ~40 arcmin patrol field. Each IFU will cover a 2.5 arcsec hexagon with ~200 milliarcseconds spaxels being fed by the GLAO system.

## F. Complementary Telescopes

### F.1. The Vera C. Rubin Observatory

Also coming on-line in the next few years is the Vera C. Rubin Observatory. This is an 8.4m Synoptic Survey Telescope located in Chile. The science goals are (1) Studying dark energy and dark matter by measuring weak gravitational lensing, baryon acoustic oscillations, and photometry of type Ia supernovae, all as a function of redshift, (2) Mapping small objects in the Solar System, particularly near-Earth asteroids and Kuiper belt objects with the expectation of increasing the number of cataloged objects by a factor of 10–100, (3) Detecting transient astronomical events including novae, supernovae, gamma-ray bursts, quasar variability, and gravitational lensing, and providing prompt event notifications to facilitate follow-up, and (4) Mapping the Milky Way. It will repeatedly image the available sky and is capable of observing 20,000 square degrees of the sky into bands every three nights using a 3.2 gigapixel camera in the wavelength range of 0.33 μm – 1.08 μm.

## G. Space Telescopes

JWST is a 6.5m Infrared telescope located at L2. Being in space it provides diffraction-limited data, down to 2 μm (limited by optics quality below this wavelength), for a number of instruments. These are NIRCam, NIRISS, NIRSpec, & MIRI. NIRCam (the Near Infrared Camera) is the JWST primary imager over the wavelength range 0.6 μm – 5 μm. It is equipped with coronagraphs enabling exoplanet observations. NIRISS (the Near Infrared Imager and Slitless Spectrograph) has a wavelength range of 0.8 μm – 5.0 μm and is applicable for exoplanet detection and characterization, as well as exoplanet transit spectroscopy. NIRSpec (the Near InfraRed Spectrograph) operates over a wavelength range of 0.6 μm – 5μm. It can obtain simultaneous spectra of more than 100 objects in a 9 square arcminute FoV and provides medium-resolution spectroscopy over a wavelength range of 1 μm – 5 μm and lower-resolution spectroscopy from 0.6 μm – 5 μm. MIRI (the Mid-Infrared Instrument) operates over a wavelength range of 5 μm – 28 μm.

The Nancy Grace Roman Telescope (NGRT) is a 2.4 telescope with the goal of settling essential questions in the areas of dark energy, exoplanets, and infrared astrophysics. It will have two instruments. The first is the Wide Field Instrument (WFI) which will provide imaging over a 0.8 x 0.4 deg FoV with a spatial sampling of 0.11”/pixel. The second is a technology demonstration, the Coronagraph, with the goal of demonstrating the technology to enable future missions to observe and characterize rocky planets in the habitable zone of their star. Once successfully demonstrated the Coronagraph could become open to the scientific community. NGRT is planned to have a 5 year mission.

Euclid is an ESA mission with the goal to map the geometry of the Universe to better understand dark matter and dark energy. It is a wide-field telescope (1.2m) with a 600-megapixel visible light camera (0.55 μm – 0.90), a near-infrared spectrometer, and photometer, with Y, J, H band spectroscopy. It will investigate the distance-redshift relationship and the evolution of cosmic structures by measuring shapes and redshifts of galaxies and clusters of galaxies out to redshifts ~2. It was launched in July 2023 and has a six year mission.